\documentclass{article}
\usepackage{amsmath}

\input{tcilatex}

\begin{document}

\title{ATLAS\ OF\ TWO-DIMENSIONAL IRREVERSIBLE\ CONSERVATIVE LAGRANGIAN\
MECHANICAL\ SYSTEMS\ WITH\ A\ SECOND\ QUADRATIC\ INTEGRAL}
\author{\textit{H. M. YEHIA} \\
\textit{Department of Mathematics, Faculty of Science,}\\
\textit{\ Mansoura University, Mansoura 35516, Egypt}\\
E-mail: hyehia@mans.edu.eg}
\maketitle

\begin{abstract}
This paper aims at the most comprehensive and systematic construction and
tabulation of mechanical systems that admit a second invariant, quadratic in
velocities, other than the Hamiltonian. The configuration space is in
general a 2D Riemannian or pseudo-Riemannian manifold and the determination
of its geometry is a part of the process of solution. Forces acting on the
system include a part derived from a scalar potential and a part derived
from a vector potential, associated with terms linear in velocities in the
Lagrangian function of the system. The last cause time-irreversibility of
the system.

We construct 41 multi-parameter integrable systems of the type described in
the title mostly on Riemannian manifolds. They are mostly new and cover all
previously known systems as special cases, corresponding to special values
of the parameters. Those include all known cases of motion of a particle in
the plane and all known cases in the dynamics of rigid body. In the last
field we introduce a new integrable case related to Steklov's case of motion
of a body in a liquid. Several new cases of motion in the plane, on the
sphere and on the pseudo-sphere or in the hyperbolic plane are found as
special cases. Prospective applications in mathematics and physics are also
pointed out.\newpage
\end{abstract}

\section{Introduction}

\subsection{Historical}

The direct method for constructing integrable time-reversible mechanical
systems was initiated by Bertrand in his papers \cite{bert0} and \cite{bert}%
. In the first one he posed and partially solved the problem of finding
forces that should act on a particle in the plane from the knowledge of an
additional integral of motion of a prescribed form. In the second paper \cite%
{bert} Bertrand formulated the problem of finding forces acting on a
particle in the plane so that the Lagrangian 
\begin{equation}
L=\frac{1}{2}(\dot{x}^{2}+\dot{y}^{2})-V(x,y)  \label{brt}
\end{equation}%
admits a first integral of a prescribed form. He obtained the equations
satisfied by the coefficients of an integral, polynomial or rational (linear
to linear) in the velocities and the components of force acting on the
system and succeeded to solve them in certain partial cases. For the case of
quadratic integral, Bertrand obtained the necessary and sufficient condition
is that $V$ satisfies the partial differential equation%
\begin{eqnarray}
&&(2axy+by+b^{\prime }x-c_{1})(V_{xx}-V_{yy})+2[a(y^{2}-x^{2})+by-b^{\prime
}x+c]V_{xy}  \nonumber \\
&&+3(2ay+b)V_{x}-3(2ax+b^{\prime })V_{y}  \nonumber \\
&=&0  \label{brte}
\end{eqnarray}%
where $a,b,...$ are constants. Darboux \cite{dar} gave the generic solution
of this equation for the potential in the form 
\begin{equation}
V=\frac{f(\alpha )-\varphi (\beta )}{\alpha ^{2}-\beta ^{2}}
\end{equation}%
where $f,\varphi $ are arbitrary functions and $\alpha ,\beta $ are the
parameters of confocal ellipses and hyperbolas in the plane. The cases of
degeneracy of elliptic coordinates to parabolic, polar and Cartesian
coordinates were considered in complete form only recently \cite{win}. The
last cases were also rediscovered more than once in later times \cite{ank}, 
\cite{thom} \cite{sen}. The same result was drawn from different points of
view in \cite{dor}, where some complex cases were pointed out (see also \cite%
{hiet} for complete description of possible real and complex cases) and \cite%
{puc}, which applied differential geometric concepts to the iso-energetic
Jacobi metric. It turns out that namely in those four cases the
Hamilton-Jacobi equation of the system (\ref{brt}) is separable \cite{stkl}, 
\cite{eisen} (see also \cite{LL}).

In the present paper we deal with a significant generalization of the system
of (\ref{brt}), described by the Lagrangian%
\begin{equation}
L=\frac{1}{2}\underset{i,j=1}{\overset{2}{\sum }}a_{ij}\dot{q}_{i}\dot{q}%
_{j}+\underset{i=1}{\overset{2}{\sum }}a_{i}\dot{q}_{i}-V  \label{0}
\end{equation}%
in which $a_{ij},a_{i},V$ depend only on the coordinates $q_{1},q_{2}$. This
system differs from (\ref{brt}) in two important aspects:

\begin{enumerate}
\item It accommodates Riemannian two-dimensional manifolds as possible
configuration spaces\footnote{%
In classical mechanics we are usually interested in systems whose kinetic
energy (equivalently, metric) is positive definite. Certain results below
involve indefinite (pseudo-Riemannian) metrics. These are included for the
sake of completeness of enumeration of all possible cases.}. This allows
application to various problems in dynamics involving motion of a rigid body
about a fixed point and motion of a particle on a fixed smooth curved
surface.

\item The terms linear in velocities account for gyroscopic forces. Those
are forces that do not do work during motion of the system. They appear
naturally in gyroscopic systems as a result of ignoring cyclic coordinates
by the Routh procedure. Linear terms may arise also when the system has some
charged components moving in a stationary magnetic field. In this context
the vector $(a_{1},a_{2})$ may be called "vector potential" while $V$ is the
scalar potential. The system (\ref{0}) is usually called a system with
velocity-dependent potential. For other circumstances see \cite{y92}.
\end{enumerate}

Birkhoff \cite{birk} raised and completely solved the problem of
constructing systems of the type (\ref{0}), which admit an integral linear
in velocities in addition to the Jacobi integral (the Hamiltonian of the
system). The first cases of irreversible systems in the Euclidean plane,
having an integral quadratic in velocities were obtained by Vandervoort \cite%
{van} and by Dorizzi et al \cite{dor85}.

The first systematic study of the full irreversible system of type (\ref{0})
for existence of a quadratic integral was done in our work \cite{y92}, where
possible integrable systems were classified according to whether the normal
form of the metric has the Euclidean form, the form of a metric on a surface
of revolution or a generic Liouville form. Several new many-parameter
families of systems were found. Most of the known systems of the same type
were either recovered or shown to be special cases of the new systems,
corresponding to certain particular values of the parameters. Some main
systems were written down explicitly, but many others were only briefly
pointed out. Some cases of integrable motions in the plane generalizing the
special plane versions of \cite{y92} were presented in \cite{ypl}. Cases of
irreversible motion on a sphere corresponding to integrable problems due to
Clebsch and Lyapunov in rigid body dynamics were obtained in \cite{ysph}. A
case of irreversible motion of a particle on a smooth ellipsoid
corresponding to Clebch's case of asymmetric body in a liquid was also
obtained in \cite{yell}. In the last two works the Lagrangian for each
motion was expressed in redundant Cartesian coordinates on the surface. They
can be simply reduced to the form (\ref{0}) by using sphero-conic
coordinates on the sphere and elliptic coordinates on the ellipsoid.

After the publication of \cite{y92} few works appeared dealing with the same
problem, but mainly investigating the case of Euclidean plane. A nearly
complete list is composed of the following papers: \cite{cont-van}, \cite%
{stew} (devoted to motion of a particle in a rotating plane) and \cite%
{mcsw-win}, \cite{puc}, \cite{puc-ros} dealing with particles moving in the
plane under the action of potential and "forces with vector potential" or
"velocity-dependent" forces. Only two works considered irreversible motions
on curved manifolds: \cite{ben} from the point of view of separation of
variables and \cite{fer} in which a classification of possible cases on
Riemannian manifolds is tried, but most of the results are very special
cases of those presented in \cite{y92}.

The aim of the present paper is to make the most exhaustive enumeration of
irreversible mechanical systems, which have an invariant quadratic in
velocities and functionally independent of the Hamiltonian of the system. We
have constructed a total of 41 different several-parameter families of those
systems. Some of those families or their special cases or degenerations were
listed in our previous works but the majority are listed for the first time.
We give briefly the very essential moments in the different methods used in
the derivation of the families and then provide for every family the minimal
information enough for its characterization. Usually, we give the Lagrangian
an the second invariant and occasionally the Gaussian curvature of the
configuration space, which we use for interpretation of suitable results as
cases of motion on a plane or a space of constant curvature. The last
section is devoted to application in the two model problems of motion of a
rigid body under various circumstances and of motion of a particle under
potential forces on a smooth fixed ellipsoid.

\subsection{Formulation of the problem}

It is well known that the system (\ref{0}) can always be referred to some
isometric coordinates $x,y$ (say), in which the Lagrangian takes the form%
\begin{equation}
L=\frac{1}{2}\Lambda _{0}(\dot{x}^{2}+\dot{y}^{2})+l_{1}\dot{x}+l_{2}\dot{y}%
-V  \label{1}
\end{equation}%
containing one coefficient in the kinetic energy term, instead of three in (%
\ref{0}). The pair of coefficients $l_{1},l_{2}$ are not unique for a given
system, since one can add the time derivative of an arbitrary function of
the position to the Lagrangian. The Lagrangian -more precisely, the
Lagrangian equations of motion- is thus characterized by the triplet of
functions $(\Lambda _{0},\Omega _{0},V),$ where $\Omega _{0}=\frac{\partial
l_{1}}{\partial y}-\frac{\partial l_{2}}{\partial x}.$

The system (\ref{1}) transforms under conformal mapping 
\begin{equation}
x+iy=z(\zeta =\xi +i\eta )  \label{1a}
\end{equation}%
to%
\begin{equation}
L=\frac{1}{2}\Lambda (\dot{\xi}^{2}+\dot{\eta}^{2})+l_{1}^{\prime }\dot{\xi}%
+l_{2}^{\prime }\dot{\eta}-V  \label{2}
\end{equation}%
where $\Lambda =\mid \frac{dz}{d\zeta }\mid ^{2}\Lambda _{0},\Omega =\mid 
\frac{dz}{d\zeta }\mid ^{2}\Omega _{0}.$ The last system admits the Jacobi
integral of motion (the Hamiltonian expressed in the state variables) 
\begin{equation}
H=\frac{1}{2}\Lambda (\dot{\xi}^{2}+\dot{\eta}^{2})+V=K\text{ (arbitrary
constant.)}  \label{3}
\end{equation}

Suppose that the system (\ref{1}) admits an integral of general quadratic
form%
\begin{equation}
I=\frac{1}{2}(b_{11}\dot{x}^{2}+2b_{12}\dot{x}\dot{y}+b_{22}\dot{y}%
^{2})+b_{1}\dot{x}+b_{2}\dot{y}+b_{0}
\end{equation}

In our previous work \cite{y92} we have shown that for a system of the type
under consideration one can always find a conformal mapping of the plane and
a time transformation of the type 
\begin{equation}
dt=\Lambda d\tau  \label{4}
\end{equation}%
such that the Lagrangian, Jacobi's integral and the complementary quadratic
integral take, with the proper use of (\ref{4}), the form \cite{y92}:

\begin{eqnarray}
L &=&\frac{1}{2}(\xi ^{\prime 2}+\eta ^{\prime 2})+P\xi ^{\prime }-Q\eta
^{\prime }+U  \label{Lf} \\
H &=&\frac{1}{2}(\xi ^{\prime 2}+\eta ^{\prime 2})-U \\
I &=&\frac{1}{2}\xi ^{\prime 2}+P\xi ^{\prime }+Q\eta ^{\prime }+R
\label{If}
\end{eqnarray}%
where primes denote differentiation with respect to $\tau ,$ 
\begin{subequations}
\begin{eqnarray}
P &=&\phi _{\eta },Q=\phi _{\xi },  \label{ex1} \\
U &=&\Lambda (K-V),  \label{ex2} \\
R &=&\int_{\eta _{0}}^{\eta }\Omega Pd\eta -(\Omega Q+U_{\xi })_{\eta =\eta
_{0}}d\xi ,  \label{ex3} \\
\Omega &=&\phi _{\xi \xi }+\phi _{\eta \eta }  \label{ex4}
\end{eqnarray}%
and $\Lambda ,\phi ,V$ satisfy the over-determined system of four equations 
\end{subequations}
\begin{subequations}
\label{EQ}
\begin{eqnarray}
\Lambda _{\xi \eta } &=&0,  \label{eq1} \\
\phi _{\eta }\Lambda _{\xi }+\phi _{\xi }\Lambda _{\eta }+2\Lambda \phi
_{\xi \eta } &=&0,  \label{eq2} \\
\phi _{\eta }V_{\xi }+\phi _{\xi }V_{\eta } &=&0,  \label{eq3} \\
\lbrack \frac{\partial ^{2}}{\partial \xi ^{2}}+\frac{\partial ^{2}}{%
\partial \eta ^{2}}](\phi _{\xi }\phi _{\eta })-(\Lambda V)_{\xi \eta } &=&0
\label{eq4}
\end{eqnarray}

The equations of motion corresponding to the Lagrangian (\ref{Lf}) acquire
the simplest form 
\end{subequations}
\begin{equation}
\xi ^{\prime \prime }+\Omega \eta ^{\prime }=\frac{\partial U}{\partial \xi }%
,\eta ^{\prime \prime }-\Omega \xi ^{\prime }=\frac{\partial U}{\partial
\eta }  \label{rem}
\end{equation}

\bigskip In the original (natural) time variable the Lagrangian and
integrals take the form 
\begin{subequations}
\label{n}
\begin{eqnarray}
L &=&\frac{1}{2}\Lambda (\dot{\xi}^{2}+\dot{\eta}^{2})+P\dot{\xi}-Q\dot{\eta}%
+K-V  \label{nl} \\
H &=&\frac{1}{2}\Lambda (\dot{\xi}^{2}+\dot{\eta}^{2})+V-K=0  \label{nj} \\
I &=&\frac{1}{2}\Lambda ^{2}\dot{\xi}^{2}+\Lambda (P\dot{\xi}+Q\dot{\eta})+R
\label{ni}
\end{eqnarray}

In the last form of the integral (\ref{ni}) the function $R$ may involve the
energy-parameter (more precisely, Jacobi's parameter) $K$ as a linear
multiplier in certain terms. This is obvious from the way of construction.
This parameter is interpreted as its numerical value on a given motion, or
may be substituted by its expression as a function of the state variables
from (\ref{nj}), resulting in an unconditional integral in the state space.

The function $\Lambda ,$ the solution of equation (\ref{eq1}) completely
characterizes the metric on the configuration space of the system (\ref{nl}%
). This gives a natural basis for a classification of all systems of the
type under consideration, depending on whether $\Lambda $ has one of three
possible forms:

1) The factor $\Lambda $ is a constant and, without loss of generality, one
can set 
\end{subequations}
\begin{equation}
\Lambda =1  \label{ty1}
\end{equation}%
In that case the configuration space has Euclidean metric. It may be
interpreted as a plane or a developable surface.

2) The factor $\Lambda $ depends only on one variable 
\begin{equation}
\Lambda =\mu (\eta )  \label{ty2}
\end{equation}%
In that case the metric on the configuration space has the structure of a
metric on a surface of revolution. This does not mean that one can always
realize this metric on a surface of revolution, but this may be done in
certain special circumstances.

3) $\Lambda $ has the generic form of a Liouville metric%
\begin{equation}
\Lambda =\lambda (\xi )-\mu (\eta )  \label{ty3}
\end{equation}

The solution of equations (\ref{eq2}-\ref{eq4}) will depend mainly on the
form of chosen solution for $\Lambda .$ The three cases will be dealt with
separately in the following sections.

\section{\protect\bigskip Time-reversible plane systems: The
Bertrand-Darboux project}

When the function $\Omega \equiv 0$ equations (\ref{rem}) become
time-reversible and one can take $\phi \equiv 0$, so that in virtue of (\ref%
{eq1}, \ref{eq4}) the Lagrangian can be written as%
\begin{equation}
L=\frac{1}{2}(\lambda (\xi )-\mu (\eta ))[\dot{\xi}^{2}+\dot{\eta}^{2}]-%
\frac{v_{1}(\xi )-v_{2}(\eta )}{(\lambda (\xi )-\mu (\eta ))}+h  \label{LS}
\end{equation}%
This is the general two-dimensional system of Liouville type, and it can be
put in the common form. \ 
\begin{equation}
L=\frac{1}{2}(\lambda -\mu )[\frac{\dot{\lambda}^{2}}{F(\lambda )}+\frac{%
\dot{\mu}^{2}}{G(\mu )}]-\frac{v_{1}(\lambda )-v_{2}(\mu )}{\lambda -\mu }+h
\label{LS1}
\end{equation}%
It involves four arbitrary functions of each of one variable each.
Separation of variables in this case is obvious.

Many cases of physical interest can be identified in the form of the
Liouville system (\ref{LS}).

\bigskip If the original configuration space of (\ref{1}) is the ordinary
Euclidean plane, then $\Lambda _{0}=1$ and we obtain a different and very
compact formulation of the Bertrand-Darboux problem presented in the
introduction. Solutions of this problem are systems of the form (\ref{LS})
with the function $z(\zeta )$ in the transformation (\ref{1a}) satisfying
the equation

\[
\Lambda \equiv \mid \frac{dz}{d\zeta }\mid ^{2}=\lambda (\xi )-\mu (\eta ) 
\]%
It was shown in \cite{y92} that this equation has the general solution 
\begin{equation}
\zeta =\int \frac{dz}{\sqrt{\alpha z^{2}+\beta z+\gamma }}  \label{r}
\end{equation}%
in which $\alpha $ is a non-negative constant and the arbitrary constants$%
\beta ,\gamma $ can be made real by a suitable choice of axes in the $z-$%
plane. The generic case of two distinct roots defines the elliptic
coordinates in the $z-$plane, reproduces Darboux's result. The degenerate
cases of two equal roots or with one or both roots coalescing with the point
at infinity define the polar, parabolic and Cartesian coordinates in the
plane. This gives a much simpler and transparent way to the completion of
the Bertrand-Darboux project than that relying on the cases of solution of
Bertrand's equation (\ref{brte}).

\section{\label{ft}The first type of irreversible systems - The metric is an
Euclidean plane one (System 1)}

We begin the classification by considering the first and simplest case, when 
$\Lambda =$const.$=1.$ It is quite straightforward to solve the system (\ref%
{EQ}) as in \cite{y92} and construct the Lagrangian in the form%
\begin{eqnarray}
L &=&\frac{1}{2}(\dot{x}^{2}+\dot{y}^{2})+\sqrt{G(\mu )}\dot{x}-\sqrt{%
F(\lambda )}\dot{y}  \nonumber \\
&&+\frac{1}{2}a(\lambda -\mu )^{3}+\frac{1}{2}(b_{1}+b_{2})(\lambda -\mu
)^{2}+K_{1}(\lambda -\mu )+h  \label{L0fk}
\end{eqnarray}%
where%
\begin{eqnarray}
x &=&\int^{\lambda }\frac{d\lambda }{\sqrt{F(\lambda )}},y=\int^{\mu }\frac{%
d\mu }{\sqrt{G(\mu )}}  \nonumber \\
F(\lambda ) &=&a\lambda ^{3}+b_{1}\lambda ^{2}+c_{1}\lambda +d_{1}, 
\nonumber \\
G(\mu ) &=&-a\mu ^{3}+b_{2}\mu ^{2}+c_{2}\mu +d_{2}  \label{fg}
\end{eqnarray}%
and $a,b_{1},b_{2},c_{1},c_{2},d_{1},d_{2},K_{1},h$ are arbitrary constants.
This is equivalent to the less explicit result provided in \cite{y92}. The
equations of motion have the form%
\begin{equation}
\ddot{x}+\Omega \dot{y}=-\frac{\partial V}{\partial x},\ddot{y}-\Omega \dot{x%
}=-\frac{\partial V}{\partial y}
\end{equation}%
where 
\begin{equation}
\Omega =\frac{1}{2}[3a(\lambda ^{2}-\mu ^{2})+2(b_{1}\lambda +b_{2}\mu
)+c_{1}+c_{2}
\end{equation}%
They admit the complementary quadratic integral%
\begin{eqnarray}
I &=&\frac{1}{2}\dot{x}^{2}+\sqrt{G(\mu )}\dot{x}+\sqrt{F(\lambda )}\dot{y} 
\nonumber \\
&&-\frac{1}{2}[a(\lambda -\mu )^{2}(2\lambda +\mu )+(c_{1}+c_{2})(\lambda
-\mu )+b_{2}(\lambda ^{2}-\mu ^{2})]-b_{1}\lambda (\lambda -\mu
)+K_{1}\lambda  \label{I0fk}
\end{eqnarray}%
The system with the Lagrangian (\ref{L0fk}) describes the motion of a
particle in the plane, with arbitrary Jacobi's constant $h$. An equivalent
of this system was pointed out in \cite{dor85} and discussed also in \cite%
{y92}. The Lagrangian can be explicitly expressed in terms of the Cartesian
coordinates $x,y$ using Weierstrass' elliptic functions of two independent
sets of invariants. Different possible real interpretations are discussed in 
\cite{y92} in terms of Jacobi's elliptic functions. Except some
degenerations, the potential $V$ and the gyroscopic functions $\Omega $ are
periodic in both $x,y$ directions. The generic system (\ref{L0fk}) may be
used as a model for the study of motion of an electrically charged particle
in the plane under the action of an electric field in the plane and a
magnetic field orthogonal to it. Both fields may be regarded as resulting
from a spatially periodic distribution of sources.

\bigskip The simplest periodic case, when elliptic functions degenerate into
trigonometric ones is characterized by%
\begin{eqnarray}
V &=&\frac{1}{2}(\alpha ^{2}+\beta ^{2})[A\cos (\alpha x)+B\cos (\beta
y)]^{2}-k[A\cos (\alpha x)+B\cos (\beta y)]  \nonumber \\
\Omega  &=&\alpha ^{2}A\cos (\alpha x)+\beta ^{2}B\cos (\beta y)  \nonumber
\\
I &=&\dot{x}^{2}-\dot{y}^{2}+4\beta B\sin (\beta y)\dot{x}+4\alpha A\sin
(\alpha x)\dot{y}-2k[A\cos (\alpha x)+B\cos (\beta y)]  \nonumber \\
&&+[A\cos (\alpha x)-B\cos (\beta y)][A(3\alpha ^{2}+\beta ^{2})\cos (\alpha
x)+B(\alpha ^{2}+3\beta ^{2})\cos (\beta y)]
\end{eqnarray}%
Even this case is a quite complicated integrable one. The following two
graphs 1, 2 show the functions $V$ and $\Omega $ on an area $4\pi \times
4\pi ,$ containing four primitive periodic cells of the plane in the special
case $\alpha =\beta =A=B=k=1.$%
\[
\FRAME{itbpF}{3.4817in}{2.8279in}{0in}{}{}{Figure}{\special{language
"Scientific Word";type "GRAPHIC";maintain-aspect-ratio TRUE;display
"USEDEF";valid_file "T";width 3.4817in;height 2.8279in;depth
0in;original-width 5.0211in;original-height 4.0733in;cropleft "0";croptop
"1";cropright "1";cropbottom "0";tempfilename
'JIK4W107.wmf';tempfile-properties "XPR";}}
\]

\begin{center}
Figure 1 :$V(x,y)$ the electric potential in the plane of motion.

\[
\FRAME{itbpF}{3.352in}{3.352in}{0in}{}{}{Figure}{\special{language
"Scientific Word";type "GRAPHIC";maintain-aspect-ratio TRUE;display
"USEDEF";valid_file "T";width 3.352in;height 3.352in;depth
0in;original-width 5.0315in;original-height 5.0315in;cropleft "0";croptop
"1";cropright "1";cropbottom "0";tempfilename
'JIK4W106.wmf';tempfile-properties "XPR";}}
\]

\bigskip Figure 2: $\Omega (x,y)$ the magnetic field orthogonal to the plane.
\end{center}

\section{\label{typ2}Second type}

The problem of constructing systems of this type was considered in \cite{y92}%
, where only few cases were given in detail. Here we give a more systematic
treatment and the final results cover a much wider hierarchy of possible
different cases, some of which were overlooked in \cite{y92}.

In fact, putting $\Lambda =\mu (\eta )$ in (\ref{EQ}) we obtain the general
solution of (\ref{eq2}) as $\varphi =\frac{M(\xi )}{\sqrt{\mu }}+N(\eta ),$
where $M,N$ are arbitrary functions. Also, from (\ref{eq3}) we deduce that $%
V $ must have the structure%
\begin{equation}
V=V(\psi ),\psi =\sqrt{\mu }M(\xi )+v(\eta )
\end{equation}%
where $v$ is related to $N$ by the condition $v^{\prime }(\eta )=-\mu
N^{\prime }(\eta ).$ In order to make the remaining equation (\ref{eq4})
more tractable we have used $\lambda =M(\xi )$ and $\mu $ as variables
instead of $\xi ,\eta $, and introduced the notation 
\begin{equation}
(\frac{d\lambda }{d\xi })^{2}=F(\lambda ),(\frac{d\mu }{d\eta })^{2}=G(\mu )
\end{equation}%
where $F$ and $G$ are functions to be determined. Now we can write 
\begin{equation}
\psi =\sqrt{\frac{\mu }{\lambda }}+v(\mu ),\qquad \varphi =\frac{1}{\sqrt{%
\lambda \mu }}-\int \frac{v^{\prime }(\mu )}{\mu }d\mu
\end{equation}%
and equation (\ref{eq4}) becomes a consistence condition which has to be
satisfied by four functions in one variable each $V(\psi ),F(\lambda ),G(\mu
)$ and $v(\lambda ).$ The last equation can be shown to have the structure%
\begin{equation}
\mu \psi _{\mu }V^{\prime \prime }(\psi )+\frac{3}{2}V^{\prime }(\psi )=S
\label{ev}
\end{equation}%
To solve this condition, i.e. to find compatible choices of the four
functions we have applied two ways that led to different findings.

\subsection{\protect\bigskip The first way: (\protect\ref{ev}) is a
differential equation}

For (\ref{ev}) to be a differential equation in $V(\psi ),$ the coefficient $%
\mu \psi _{\mu }$ and the function $S$ must be functions of the single
variable $\psi .$ The first implies that $v(\mu )=k\sqrt{\mu }.$ Turning to
the function $S$ we have found that $k=0$ and also 
\begin{eqnarray}
F(\lambda ) &=&\lambda ^{2}(a_{3}\lambda ^{3}+a_{2}\lambda ^{2}+a_{1}\lambda
+a_{0})  \nonumber \\
G(\mu ) &=&\mu ^{2}(b_{3}\mu ^{3}+b_{2}\mu ^{2}+b_{1}\mu -a_{0})
\end{eqnarray}%
Returning to equation (\ref{ev}) we find 
\begin{equation}
V(\psi )=-\frac{1}{8}(b_{3}\psi ^{2}+\frac{a_{3}}{\psi ^{4}})+\frac{A}{\psi
^{2}},\qquad \psi =\sqrt{\frac{\mu }{\lambda }}
\end{equation}

By this information we are now able to construct the system we are looking
for. We call it System 2. It has the Lagrangian%
\begin{eqnarray}
L &=&\frac{1}{2}\mu \lbrack \frac{\dot{\lambda}^{2}}{F(\lambda )}+\frac{\dot{%
\mu}^{2}}{G(\mu )}]  \nonumber \\
&&+\frac{1}{2}J\sqrt{\frac{F(\lambda )G(\mu )}{\lambda ^{3}\mu ^{3}}}[\frac{%
\lambda \dot{\lambda}}{F(\lambda )}-\frac{\mu \dot{\mu}}{F(\mu )}]  \nonumber
\\
&&-\frac{A\lambda }{\mu }+J^{2}\frac{1}{8}(a_{3}\frac{\lambda ^{2}}{\mu ^{2}}%
+b_{3}\frac{\mu }{\lambda })  \label{L1}
\end{eqnarray}

\subsection{The second way: (\protect\ref{ev}) is a consistency condition:}

We deal with (\ref{ev}) assuming for $V(\psi )$ a preassigned form like a
polynomial and try to find compatible combinations of coefficients that help
separate equations for $F,G$ and $v.$ The result found in this way can be
formulated as in the following

\begin{theorem}
\bigskip
\end{theorem}

The system with the Lagrangian%
\begin{eqnarray}
L &=&\frac{1}{2}[\frac{\mu \dot{\lambda}^{2}}{a_{0}+a_{1}\lambda
-c_{0}\lambda ^{2}}+\frac{\dot{\mu}^{2}}{4\mu (c_{3}\mu ^{3}+c_{2}\mu
^{2}+c_{1}\mu +c_{0})}]  \nonumber \\
&&-[\frac{J\lambda }{\sqrt{\mu }}+2v^{\prime }(\mu )]\frac{\sqrt{c_{3}\mu
^{3}+c_{2}\mu ^{2}+c_{1}\mu +c_{0}}}{\sqrt{a_{0}+a_{1}\lambda -c_{0}\lambda
^{2}}}\dot{\lambda}  \nonumber \\
&&-\frac{J\sqrt{a_{0}+a_{1}\lambda -c_{0}\lambda ^{2}}\dot{\mu}}{2\mu ^{3/2}%
\sqrt{c_{3}\mu ^{3}+c_{2}\mu ^{2}+c_{1}\mu +c_{0}}}  \nonumber \\
&&-b[J\sqrt{\mu }\lambda +v(\mu )]+\frac{1}{2}c_{3}[J\sqrt{\mu }\lambda
+v(\mu )]^{2}  \label{Lr}
\end{eqnarray}%
where $c_{j},a_{i},b,J$ are arbitrary constants and $v(\mu )$ is the
solution of the linear differential equation%
\begin{gather}
8(c_{3}\mu ^{3}+c_{2}\mu ^{2}+c_{1}\mu +c_{0})v^{\prime \prime \prime }(\mu
)+12(3c_{3}\mu ^{2}+2c_{2}\mu +c_{1})v^{\prime \prime }(\mu )  \nonumber \\
+6(3c_{3}\mu +c_{2})v^{\prime }(\mu )-3c_{3}v(\mu )=-3b-\frac{3Ja_{1}}{2\mu
^{5/2}}  \label{e2r}
\end{gather}%
admits the second quadratic integral%
\begin{eqnarray}
I &=&\frac{1}{2}\frac{\mu ^{2}\dot{\lambda}^{2}}{a_{0}+a_{1}\lambda
-c_{0}\lambda ^{2}}  \nonumber \\
&&-\mu \lbrack \frac{J\lambda }{\sqrt{\mu }}+2v^{\prime }(\mu )]\frac{\sqrt{%
c_{3}\mu ^{3}+c_{2}\mu ^{2}+c_{1}\mu +c_{0}}}{\sqrt{a_{0}+a_{1}\lambda
-c_{0}\lambda ^{2}}}\dot{\lambda}  \nonumber \\
&&+\frac{J\mu \sqrt{a_{0}+a_{1}\lambda -c_{0}\lambda ^{2}}\dot{\mu}}{2\mu
^{3/2}\sqrt{c_{3}\mu ^{3}+c_{2}\mu ^{2}+c_{1}\mu +c_{0}}}  \nonumber \\
&&+\frac{1}{2}J^{2}\lambda \lbrack \lambda \mu (c_{3}\mu +c_{2})-\frac{a_{1}%
}{\mu }]-\frac{1}{2}Ja_{1}\int \frac{v^{\prime }(\mu )}{\mu ^{3/2}}d\mu 
\nonumber \\
&&+J\lambda \sqrt{\mu }\{4(c_{3}\mu ^{3}+c_{2}\mu ^{2}+c_{1}\mu
+c_{0})v^{\prime \prime }(\mu )  \nonumber \\
&&+2(3c_{3}\mu ^{2}+2c_{2}\mu +c_{1})v^{\prime }(\mu )-c_{3}\mu v(\mu )+b\mu
\}  \nonumber \\
&&+2v^{\prime }(\mu )^{2}(c_{3}\mu ^{3}+c_{2}\mu ^{2}+c_{1}\mu +c_{0})
\label{Ir}
\end{eqnarray}

When $c_{0}\neq 0$ the above theorem gives the same result as in \cite{y92},
as it is possible in this case to eliminate $a_{1}$ by a suitable shift in
the variable $\lambda .$ But when $c_{0}=0$ the presence of $a_{1}$ is
significant and leads to seven new cases in dependence on the combination of
roots of $G(\mu ).$ The complete hierarchy is given in the next subsection.

\subsubsection{Classification}

For the complete determination of the Lagrangian (\ref{Lr}) and the integral
(\ref{Ir}) we need only the solutions of (\ref{e2r}) for $v(\mu )$. This
turned out to be depending on the distribution of roots of the polynomial $%
G(\mu )/\mu ^{2}=c_{3}\mu ^{3}+c_{2}\mu ^{2}+c_{1}\mu +c_{0},$ which we
denote by $\mu _{1},\mu _{2},\mu _{3}.$ We now proceed to enumerate the
fourteen possible cases. For each case we give its conditions and the
corresponding solution$v(\mu )$.

\begin{tabular}{|l|ll|l|l|}
\hline
& \multicolumn{2}{|l}{\qquad Conditions on coefficients} & Roots & $v(\mu )$
\\ \hline
3 & \multicolumn{1}{|l|}{} & \multicolumn{1}{l|}{} & $\mu _{3}\neq \mu
_{2}\neq \mu _{1}\neq 0$ & $%
\begin{tabular}{l}
$\frac{b}{c_{3}}+K_{1}\sqrt{\mu -\mu _{1}}$ \\ 
$+K_{2}\sqrt{\mu -\mu _{2}}+K_{3}\sqrt{\mu -\mu _{3}}$%
\end{tabular}%
$ \\ \cline{1-1}\cline{4-5}
4 & \multicolumn{1}{|l|}{} & \multicolumn{1}{l|}{$c_{0}\neq 0$} & $\mu
_{3}=\mu _{2}\neq \mu _{1}\neq 0$ & 
\begin{tabular}{l}
$\frac{b}{c_{3}}+K_{1}\sqrt{\mu -\mu _{1}}+K_{2}\sqrt{\mu -\mu _{2}}$ \\ 
$+K_{3}/\sqrt{\mu -\mu _{2}}$%
\end{tabular}
\\ \cline{1-1}\cline{4-5}
5 & \multicolumn{1}{|l|}{} & \multicolumn{1}{l|}{} & $\mu _{3}=\mu _{2}=\mu
_{1}\neq 0$ & 
\begin{tabular}{l}
$\frac{b}{c_{3}}+K_{1}\sqrt{\mu -\mu _{1}}+K_{2}/\sqrt{\mu -\mu _{1}}$ \\ 
$+K_{3}/(\mu -\mu _{1})^{3/2}$%
\end{tabular}
\\ \cline{1-1}\cline{3-5}
6 & \multicolumn{1}{|l|}{$c_{3}\neq 0$} & $c_{0}=0,c_{1}\neq 0$ & $\mu
_{3}\neq \mu _{2}\neq \mu _{1}=0$ & 
\begin{tabular}{l}
$\frac{b}{c_{3}}+\frac{Ja_{1}}{4c_{3}\mu _{2}\mu _{3}\sqrt{\mu }}+K_{2}\sqrt{%
\mu -\mu _{2}}$ \\ 
$+K_{3}\sqrt{\mu -\mu _{3}}$%
\end{tabular}
\\ \cline{1-1}\cline{3-5}
7 & \multicolumn{1}{|l|}{} & $c_{0}=0,c_{1}\neq 0$ & $\mu _{3}=\mu _{2}\neq
\mu _{1}=0$ & $%
\begin{tabular}{l}
$\frac{b}{c_{3}}+\frac{Ja_{1}}{4c_{3}\mu _{2}\mu _{3}\sqrt{\mu }}+K_{2}\sqrt{%
\mu -\mu _{2}}$ \\ 
$+\frac{K_{3}}{\sqrt{\mu -\mu _{2}}}$%
\end{tabular}%
$ \\ \cline{1-1}\cline{3-5}
8 & \multicolumn{1}{|l|}{} & $c_{0}=c_{1}=0,c_{2}\neq 0$ & $\mu _{3}\neq \mu
_{2}=\mu _{1}=0$ & $\frac{b}{c_{3}}+\frac{K_{1}}{\sqrt{\mu }}-\frac{Ja_{1}}{%
16c_{3}\mu _{3}\mu ^{3/2}}+K_{3}\sqrt{\mu -\mu _{3}}$ \\ 
\cline{1-1}\cline{3-5}
9 & \multicolumn{1}{|l|}{} & $c_{0}=c_{1}=c_{2}=0$ & $\mu _{3}=\mu _{2}=\mu
_{1}=0$ & $\frac{b}{c_{3}}+\frac{K_{1}}{\sqrt{\mu }}+\frac{K_{2}}{\mu ^{3/2}}%
+\frac{Ja_{1}}{32c_{3}\mu ^{5/2}}$ \\ \cline{1-1}\cline{1-4}\cline{3-5}
10 & \multicolumn{1}{|l|}{$c_{3}=0,c_{2}\neq 0$} & \multicolumn{1}{|l|}{$%
c_{0}\neq 0$} & $\mu _{2}\neq \mu _{1}\neq 0$ & $-\frac{b\mu }{2c_{2}}+K_{1}%
\sqrt{\mu -\mu _{1}}+K_{2}\sqrt{\mu -\mu _{2}}$ \\ 
\cline{1-1}\cline{3-4}\cline{3-5}
11 & \multicolumn{1}{|l|}{} &  & $\mu _{2}=\mu _{1}\neq 0$ & $-\frac{b\mu }{%
2c_{2}}+K_{1}\sqrt{\mu -\mu _{1}}+K_{2}/\sqrt{\mu -\mu _{1}}$ \\ 
\cline{1-1}\cline{4-5}
12 & \multicolumn{1}{|l|}{} & $c_{0}=0$ & $\mu _{2}\neq \mu _{1}=0$ & $-%
\frac{b\mu }{2c_{2}}-\frac{Ja_{1}}{4c_{2}\mu _{2}\sqrt{\mu }}+K_{2}\sqrt{\mu
-\mu _{2}}$ \\ \cline{1-1}\cline{3-5}
13 & \multicolumn{1}{|l|}{} & $c_{0}=c_{1}=0$ & $\mu _{2}=\mu _{1}=0$ & $-%
\frac{b\mu }{2c_{2}}+\frac{K_{1}}{\sqrt{\mu }}+\frac{Ja_{1}}{16c_{2}\mu
^{3/2}}$ \\ \hline
14 & \multicolumn{1}{|l|}{%
\begin{tabular}{l}
$c_{3}=c_{2}=0$ \\ 
$c_{1}\neq 0$%
\end{tabular}%
} & $c_{0}\neq 0$ &  & $-\frac{b\mu ^{2}}{8c_{1}}+K_{1}\mu +K_{2}\sqrt{%
c_{1}\mu +c_{0}}$ \\ \cline{1-1}\cline{3-5}
15 & \multicolumn{1}{|l|}{} & $c_{0}=0$ &  & $-\frac{b\mu ^{2}}{8c_{1}}%
+K_{1}\mu +\frac{Ja_{1}}{4c_{1}\sqrt{\mu }}$ \\ \hline
16 & $%
\begin{tabular}{l}
$c_{3}=c_{2}$ \\ 
$=c_{1}=0$%
\end{tabular}%
$ & \multicolumn{1}{|l|}{$c_{0}\neq 0$} &  & $-\frac{b\mu ^{3}}{16c_{0}}%
+K_{2}\mu ^{2}+K_{1}\mu $ \\ \hline
\end{tabular}

\begin{center}
Table I
\end{center}

The Gaussian curvature of the configuration space of (\ref{Lr}) is%
\begin{eqnarray}
\kappa &=&-(3c_{3}\mu ^{2}+2c_{2}\mu +c_{1})  \nonumber \\
&=&-F^{\prime }(\mu )  \label{gc}
\end{eqnarray}%
The variable $\mu $ should be restricted to an interval $[a,b]$ (say), where 
$a,b\in \{-\infty ,0,\infty ,\mu _{1},\mu _{2},\mu _{3}\}$ on which $F(\mu
)\geq 0.$ The Gaussian curvature $\kappa $ on the configuration manifold is
of variable sign in general. But it is possible to have manifolds of
curvature of definite sign. An example is when one of the two bounds is
taken to be the point at infinity. We note also that in order that the
Lagrangian was real, the coefficients $K_{i}$ should be taken real,
imaginary or including a complex conjugate pair, depending on the nature and
relative order of the roots of $F.$ From (\ref{gc}) we also conclude that
only case 16 lives on a flat manifold and only cases 14, 15 has a manifold
of constant non-zero curvature. In the last two cases the system describes
either:

\begin{enumerate}
\item a motion on a sphere (when $c_{1}<0$), which can be recognized as the
axisymmetric version of Clebsch's integrable case of motion in a liquid of a
body with a spherical inertia tensor. The full Clebsch case is obtained as a
special version of another new integrable system in \ref{cleb}.

\item or a motion on a pseudo-sphere or in the hyperbolic plane $H^{2}$ when 
$c_{1}>0.$
\end{enumerate}

\subsubsection{Example: A case of motion on a Riemannian manifold of
positive Gaussian curvature}

In order that the Lagrangian (\ref{Lr}) represents a real system, the
kinetic energy should be positive definite. In case 6 this implies that the
variable $\mu $ should be confined to an interval of the non-negative half
of the $\mu $-axis on which $c_{2}\mu ^{2}+c_{1}\mu +c_{0}\geq 0.$ Various
possible choices of the interval lead to different configuration manifolds.
We give an example of motion on a convex manifold:

In case 10 of table I, let $c_{0}(=c_{2}\mu _{1}\mu
_{2})=1,a_{1}=0,a_{0}=\alpha ^{2}$ ,$0\leq \mu \leq \mu _{1}\leq \mu _{2},$
The substitution $\lambda =\alpha \cos \varphi ,\mu =\mu _{1}\limfunc{sn}%
^{2}u,K_{1}=ik_{1},K_{2}=ik_{2}$ ($\sqrt{\mu -\mu _{1}}=i\sqrt{\mu _{1}}%
\limfunc{cn}u,$ $\sqrt{\mu -\mu _{2}}=i\sqrt{\mu _{2}}\limfunc{dn}u$), where
the modulus of the elliptic functions $k=\sqrt{\frac{\mu _{1}}{\mu _{2}}}$
transforms the Lagrangian to the form%
\begin{eqnarray}
L &=&\frac{1}{2}\mu _{1}(\dot{u}^{2}+\limfunc{sn}\nolimits^{2}u\text{ }\dot{%
\varphi}^{2})+\frac{\alpha \sin \varphi }{\sqrt{c_{2}\mu _{2}}\mu _{1}%
\limfunc{sn}^{2}u}\dot{u}  \nonumber \\
&&+\sqrt{c_{2}}[k_{1}\sqrt{\mu _{2}}\limfunc{dn}u+k_{2}\sqrt{\mu _{1}}%
\limfunc{cn}u+\sqrt{\mu _{1}\mu _{2}}(\frac{b}{c_{2}}-\frac{\alpha \cos
\varphi }{\sqrt{\mu _{1}}\limfunc{sn}u})\limfunc{cn}u\limfunc{dn}u]\dot{x} 
\nonumber \\
&&-b[\alpha \sqrt{\mu _{1}}\limfunc{sn}u\text{ }\cos \varphi -\frac{b\mu _{1}%
}{2c_{2}}\limfunc{sn}\nolimits^{2}u-k_{1}\sqrt{\mu _{1}}\limfunc{cn}u-k_{2}%
\sqrt{\mu _{2}}\limfunc{dn}u]
\end{eqnarray}

The Gaussian curvature of the metric of this system is strictly positive%
\begin{equation}
\kappa =c_{2}(\mu _{1}\limfunc{cn}\nolimits^{2}u+\mu _{2}\limfunc{dn}%
\nolimits^{2}u)\geq c_{2}(\mu _{2}-\mu _{1})>0
\end{equation}%
It is possible to realize the Riemannian metric of the last system on a
closed surface of revolution. In fact, if $(\rho (u),\varphi ,z$ $(u))$ is a
current point of the surface in cylindrical coordinates, then we have%
\begin{equation}
d\rho ^{2}+dz^{2}+\rho ^{2}d\varphi ^{2}=\mu _{1}(du^{2}+\limfunc{sn}%
\nolimits^{2}u\text{ }d\varphi ^{2})
\end{equation}%
Comparing both sides we find that%
\begin{equation}
\rho =\sqrt{\mu _{1}}\limfunc{sn}u  \label{rho}
\end{equation}%
and hence we get an equation for $z$ as%
\[
\mu _{1}\limfunc{cn}\nolimits^{2}u\limfunc{dn}\nolimits^{2}u+(\frac{dz}{du}%
)^{2}=\mu _{1}
\]%
which, on separation and integration, gives 
\begin{equation}
z=\sqrt{\mu _{1}}\int_{0}^{u}\sqrt{1+k^{2}-k^{2}\limfunc{sn}\nolimits^{2}u}%
\limfunc{sn}udu  \label{z}
\end{equation}%
Noting that this is a periodic function with period $4K(k),$ we deduce that
the surface thus constructed is a closed one. This family of surfaces of
revolution depends on two parameters $\mu _{1}$ and $k.$ The first can be
absorbed by a suitable scaling, but $k$ changes the form of the surface.
Figure 3 shows this surface for $k=\frac{1}{\sqrt{2}}.$%
\[
\FRAME{itbpF}{2.0289in}{2.866in}{0in}{}{}{Figure}{\special{language
"Scientific Word";type "GRAPHIC";maintain-aspect-ratio TRUE;display
"USEDEF";valid_file "T";width 2.0289in;height 2.866in;depth
0in;original-width 1.9899in;original-height 2.8228in;cropleft "0";croptop
"1";cropright "1";cropbottom "0";tempfilename
'JIK4W105.wmf';tempfile-properties "XPR";}}
\]

\begin{center}
Figure 3: The surface of revolution described by (\ref{rho}) and(\ref{z})
for $k=\frac{1}{\sqrt{2}}$
\end{center}

\section{The third type - Systems with a Liouville-type metric}

\subsection{General considerations}

Now we consider the solution of the system of equations (\ref{EQ}) in the
generic case when $\lambda ^{\prime }(\xi )\mu ^{\prime }(\eta )\neq 0.$ It
will be easier in this case to use $\lambda ,\mu $ as variables. Let 
\begin{equation}
\lambda ^{\prime 2}(\xi )=F(\lambda ),\mu ^{\prime 2}(\eta )=G(\mu )
\end{equation}%
so that 
\begin{equation}
\xi =\int \frac{d\lambda }{\sqrt{F(\lambda )}},\eta =\int \frac{d\mu }{\sqrt{%
G(\mu )}}
\end{equation}%
Equations (\ref{eq2}-\ref{eq4}) can be written as 
\begin{subequations}
\label{fv}
\begin{gather}
\phi _{\lambda }-\phi _{\mu }-2(\lambda -\mu )\phi _{\lambda \mu }=0
\label{fv1} \\
\phi _{\mu }V_{\lambda }+\phi _{\lambda }V_{\mu }=0  \label{fv2} \\
\lbrack (\sqrt{F(\lambda )}\frac{\partial }{\partial \lambda })^{2}+(\sqrt{%
G(\mu )}\frac{\partial }{\partial \mu })^{2}][\sqrt{F(\lambda )G(\mu )}\phi
_{\lambda }\phi _{\mu }]  \nonumber \\
-\sqrt{F(\lambda )G(\mu )}[(\lambda -\mu )V]_{\lambda \mu }=0  \label{fv3}
\end{gather}%
One of those equations can be readily integrated. In fact, the equation of
the characteristics of (\ref{fv2}) for $V$ is 
\end{subequations}
\begin{equation}
\phi _{\lambda }d\lambda -\phi _{\mu }d\mu =0
\end{equation}%
Multiplying this equation by $(\lambda -\mu )$ one can verify that the
resulting expression is a total differential, in virtue of (\ref{fv1}), so
that one can write%
\begin{equation}
d\psi =(\lambda -\mu )(\phi _{\lambda }d\lambda -\phi _{\mu }d\mu )
\end{equation}%
Thus, the potential $V$ can be expressed as $V=V(\psi ),$ where $\psi $ is
related to $\phi $ by the relations%
\begin{equation}
\psi _{\lambda }=(\lambda -\mu )\phi _{\lambda },\qquad \psi _{\mu
}=-(\lambda -\mu )\phi _{\mu }  \label{ba}
\end{equation}%
Since only the derivatives of $\phi $ enter in (\ref{fv1}) and (\ref{fv3}),
one can completely eliminate $\phi $ from those equations. This gives for $%
\psi $ the equation%
\begin{equation}
\psi _{\lambda }-\psi _{\mu }+2(\lambda -\mu )\psi _{\lambda \mu }=0
\label{re1}
\end{equation}%
\begin{eqnarray}
&&[(\sqrt{F(\lambda )}\frac{\partial }{\partial \lambda })^{2}+(\sqrt{G(\mu )%
}\frac{\partial }{\partial \mu })^{2}](\sqrt{F(\lambda )G(\mu )}\frac{\psi
_{\lambda }\psi _{\mu }}{(\lambda -\mu )^{2}})  \nonumber \\
&&-(\lambda -\mu )\sqrt{F(\lambda )G(\mu )}[V^{^{\prime \prime }}(\psi )\psi
_{\lambda }\psi _{\mu }+3V^{\prime }(\psi )\psi _{\lambda \mu }]  \nonumber
\\
&=&0  \label{re2}
\end{eqnarray}

The Lagrangian describing the system (in the fictitious time) may now be
written as%
\begin{eqnarray}
L &=&\frac{1}{2}[\frac{\lambda ^{\prime 2}}{F(\lambda )}+\frac{\mu ^{\prime
2}}{G(\mu )}]  \nonumber \\
&&-\frac{1}{(\lambda -\mu )}\sqrt{F(\lambda )G(\mu )}[\frac{\psi _{\mu }}{%
F(\lambda )}\lambda ^{\prime }+\frac{\psi _{\lambda }}{G(\mu )}\mu ^{\prime
}]  \nonumber \\
&&+(\lambda -\mu )[K-V(\psi )]  \label{L3t}
\end{eqnarray}%
and in the natural time it takes the form%
\begin{eqnarray}
L &=&\frac{1}{2}(\lambda -\mu )[\frac{\dot{\lambda}^{2}}{F(\lambda )}+\frac{%
\dot{\mu}^{2}}{G(\mu )}]  \nonumber \\
&&-\frac{1}{(\lambda -\mu )}\sqrt{F(\lambda )G(\mu )}[\frac{\psi _{\mu }}{%
F(\lambda )}\dot{\lambda}+\frac{\psi _{\lambda }}{G(\mu )}\dot{\mu}] 
\nonumber \\
&&+K-V(\psi )  \label{Lrt}
\end{eqnarray}%
In certain circumstances we need also the expression%
\begin{equation}
\Omega =F(\lambda )\frac{\partial }{\partial \lambda }(\frac{\psi _{\lambda }%
}{\lambda -\mu })-G(\mu )\frac{\partial }{\partial \mu }(\frac{\psi _{\mu }}{%
\lambda -\mu })+\frac{F^{\prime }(\lambda )\psi _{\lambda }-G^{\prime }(\mu
)\psi _{\mu }}{2(\lambda -\mu )}  \label{oms}
\end{equation}

The process of constructing time-irreversible mechanical systems with a
quadratic integral is now reduced to the simultaneous solution of the pair
of equations (\ref{re1}) and (\ref{re2}), which involve the function $\psi
(\lambda ,\mu )$ together with three functions of one variable each: $%
F(\lambda ),G(\mu )$ and$V(\psi ).$ It is to be noted also that the three
one-variable functions occur only linearly in the last equation. This
becomes obvious if we write down this equation in the expanded form%
\begin{eqnarray}
&&[F^{\prime \prime }(\lambda )+G^{\prime \prime }(\mu )]\frac{\psi
_{\lambda }\psi _{\mu }}{(\lambda -\mu )^{2}}  \nonumber \\
&&+3\{[\frac{\psi _{\lambda }\psi _{\mu }}{(\lambda -\mu )^{2}}]_{\lambda
}F^{\prime }(\lambda )+[\frac{\psi _{\lambda }\psi _{\mu }}{(\lambda -\mu
)^{2}}]_{\mu }G^{\prime }(\mu )\}  \nonumber \\
&&+2\{[\frac{\psi _{\lambda }\psi _{\mu }}{(\lambda -\mu )^{2}}]_{\lambda
\lambda }F(\lambda )-[\frac{\psi _{\lambda }\psi _{\mu }}{(\lambda -\mu )^{2}%
}]_{\mu \mu }G(\mu )\}  \nonumber \\
&&+2(\lambda -\mu )[V^{^{\prime \prime }}(\psi )\psi _{\lambda }\psi _{\mu
}+3V^{\prime }(\psi )\psi _{\lambda \mu }]  \nonumber \\
&=&0  \label{re2a}
\end{eqnarray}

\subsection{Solving the equation for $\protect\psi $}

Equation (\ref{re1}) and (\ref{fv1}) were given an interesting
interpretation in \cite{y92}. We provide it here and use it in a more
systematic way to construct simple solutions of that equation, future
candidates of satisfying (\ref{re2}). In fact, if we introduce new variables 
$\rho =\lambda -\mu $ and $z=i(\lambda +\mu )$ then (\ref{fv1}), (\ref{re1})
and (\ref{ba}) reduce to 
\begin{subequations}
\label{sto}
\begin{eqnarray}
\phi _{\rho \rho }+\frac{1}{\rho }\phi _{\rho }+\phi _{zz} &=&0  \label{sto1}
\\
\psi _{\rho \rho }-\frac{1}{\rho }\psi _{\rho }+\psi _{zz} &=&0  \label{sto2}
\\
(-i\psi )_{z} &=&-\rho \phi _{\rho },\qquad (-i\psi )_{\rho }=\rho \phi _{z}
\label{sto3}
\end{eqnarray}%
Imagine $z$ as the axis of cylindrical coordinates in a three-dimensional
complex space and $\rho $ is the radial distance of the current point from
that axis. Equation (\ref{sto1}) is just Laplace's equation satisfied by $%
\phi -$ the velocity potential of a virtual flow of an ideal incompressible
fluid, symmetric around the $z$ axis, (\ref{sto2}) is the equation for $\psi
-$ Stokes stream function of that flow and (\ref{sto3}) are the well-known
relations between the two functions (see e.g. \cite{lamb}). In \cite{y92} we
have utilized some known axisymmetric hydrodynamic flows to construct
certain solutions of (\ref{re1}).

Here we provide a systematic way to construct solutions of (\ref{re1}) and (%
\ref{fv1}). It is well known that (\ref{sto1}) admits the spherically
symmetric solution $\phi =\frac{1}{r}=\frac{1}{\sqrt{z^{2}+\rho ^{2}}}$
representing a source at the origin. For a source on the $z-$ axis at a
distance $z_{0}$ from the origin the velocity potential becomes 
\end{subequations}
\begin{eqnarray}
\phi &=&\frac{1}{\sqrt{z_{0}^{2}-2z_{0}z+z^{2}+\rho ^{2}}}  \label{p0} \\
&=&\frac{1}{\sqrt{z_{0}^{2}-2iz_{0}(\lambda +\mu )-4\lambda \mu }}
\end{eqnarray}%
This function has two types of expansions: 
\begin{equation}
\phi =\sum\limits_{n=0}^{\infty }\frac{k_{n}}{z_{0}^{n+1}}\phi _{n}
\end{equation}%
for large $z_{0}$ and 
\begin{equation}
\phi =\sum\limits_{n=0}^{\infty }k_{n}^{\prime }z_{0}^{n}\frac{\phi _{n}}{%
(\lambda \mu )^{n+1/2}}
\end{equation}%
for small $z_{0},$ where $\phi _{n}$ is a homogeneous polynomial of degree $%
n $ in $\lambda ,\mu $ and $\{k_{n}\},\{k_{n}^{\prime }\}$ are numerical
(real or imaginary) coefficients, which can be used to give the polynomials $%
\phi _{n}$ the most convenient form. Since the expansion is valid for
arbitrary (large or small) values of $z_{0},$ we conclude that the two
infinite sequences of functions $\{\phi _{n}\},\{\frac{\phi _{n}}{(\lambda
\mu )^{n+1/2}}\}$ are solutions of equation (\ref{fv1}). It is noteworthy to
mention that those functions correspond to the familiar sequences of
axisymmetric solutions $r^{n}P_{n}(\cos \theta ),r^{-(n+1)}P_{n}(\cos \theta
)$ of Laplace's equation in spherical coordinates.

We now follow a parallel line for solving equation (\ref{re1}). The
spherically symmetric solution of Stokes equation (\ref{sto2}) is easily
seen to be $\psi =r=\sqrt{z^{2}+\rho ^{2}}.$ Displacing the source as
indicated above we get Stokes function 
\begin{equation}
\psi =\sqrt{z_{0}^{2}-2iz_{0}(\lambda +\mu )-4\lambda \mu }
\end{equation}%
For this function we obtain the two expansions%
\begin{equation}
\psi =\sum\limits_{n=0}^{\infty }\frac{K_{n}}{z_{0}^{n}}\psi _{n}
\end{equation}%
and 
\begin{equation}
\psi =\sqrt{\lambda \mu }\sum\limits_{n=0}^{\infty }K_{n}^{\prime }z_{0}^{n}%
\frac{\psi _{n}}{(\lambda \mu )^{n}}
\end{equation}%
thus giving the two infinite sequences of solutions of Stokes equation \ref%
{sto2}: $\{\psi _{n},n=0,...,\infty \},\{\tilde{\psi}_{n}=\frac{\sqrt{%
\lambda \mu }\psi _{n-1}}{(\lambda \mu )^{n-1}},n=1,...,\infty \}$. The
first few non-constant $\psi _{n}$'s are 
\begin{eqnarray}
\psi _{0} &=&1  \nonumber \\
\psi _{1} &=&\lambda +\mu  \nonumber \\
\psi _{2} &=&(\lambda -\mu )^{2}  \nonumber \\
\psi _{3} &=&(\lambda -\mu )^{2}(\lambda +\mu )  \nonumber \\
\psi _{4} &=&(\lambda -\mu )^{2}(5\lambda ^{2}+5\mu ^{2}+6\lambda \mu ) 
\nonumber \\
\psi _{5} &=&(\lambda -\mu )^{2}(\lambda +\mu )(7\lambda ^{2}+7\mu
^{2}+2\lambda \mu )  \label{pss}
\end{eqnarray}%
The first few of the second sequence $\{\tilde{\psi}_{n}\}$ are%
\begin{equation}
\sqrt{\lambda \mu },\frac{\lambda +\mu }{\sqrt{\lambda \mu }},\frac{(\lambda
-\mu )^{2}}{(\lambda \mu )^{3/2}},\frac{(\lambda -\mu )^{2}(\lambda +\mu )}{%
(\lambda \mu )^{5/2}},\frac{(\lambda -\mu )^{2}(5\lambda ^{2}+5\mu
^{2}+6\lambda \mu )}{(\lambda \mu )^{7/2}},\cdot \cdot \cdot  \label{psr}
\end{equation}%
Finally, we note that equation (\ref{re1}) is invariant with respect to
shifting $(\lambda ,\mu )$ to $(\lambda -\nu ,\mu -\nu )$ for all constant $%
\nu $, so that the sequences of solutions (\ref{pss},\ref{psr}) remain valid
after this shift. We shall use this freedom in certain circumstances.

\subsection{Adapting a solution}

A possible way of solving (\ref{re1}-\ref{re2a}) is choosing a solution $%
\psi (\lambda ,\mu )$ for the linear equation (\ref{re1}) and inserting this
solution in (\ref{re2a}), which will serve as a consistency condition. A
working choice $\psi (\lambda ,\mu )$ should allow the existence of $F,G,V$,
so that this equation is satisfied. It is natural that some solutions of (%
\ref{re1}) will be accepted in (\ref{re2a}), while others will be excluded.
Till now, we have not been able to separate any part of the last equation to
operate independently on one or two functions. It remains a matter of guess
and trial to find possible cases. It is not possible in this way to claim
that this way will lead to construction of all possible cases of the type
under consideration. However, it has lead to several rich families of
systems in one of the ways described below.

\subsubsection{\label{fp}The first procedure:}

In general, equation (\ref{re2a}), which may be written in the form%
\begin{equation}
V^{^{\prime \prime }}(\psi )+\frac{\psi _{\lambda \mu }}{\psi _{\lambda
}\psi _{\mu }}3V^{\prime }(\psi )=N(\lambda ,\mu )  \label{Ve}
\end{equation}%
where%
\begin{eqnarray}
N=\{ &&[F^{\prime \prime }(\lambda )+G^{\prime \prime }(\mu )]\frac{\psi
_{\lambda }\psi _{\mu }}{(\lambda -\mu )^{2}}  \nonumber \\
&&+3\{[\frac{\psi _{\lambda }\psi _{\mu }}{(\lambda -\mu )^{2}}]_{\lambda
}F^{\prime }(\lambda )+[\frac{\psi _{\lambda }\psi _{\mu }}{(\lambda -\mu
)^{2}}]_{\mu }G^{\prime }(\mu )\}  \nonumber \\
&&+2\{[\frac{\psi _{\lambda }\psi _{\mu }}{(\lambda -\mu )^{2}}]_{\lambda
\lambda }F(\lambda )-[\frac{\psi _{\lambda }\psi _{\mu }}{(\lambda -\mu )^{2}%
}]_{\mu \mu }G(\mu )\}\}/[2(\lambda -\mu )\psi _{\lambda }\psi _{\mu }]
\label{Ne}
\end{eqnarray}%
is a consistency condition which should be satisfied by the four functions $%
F(\lambda ),G(\mu ),V(\psi ),$ and $\psi (\lambda ,\mu ),$ of which only $%
\psi $ is subject to another equation, namely, (\ref{re1}). An important
situation is met when (\ref{Ve}) becomes a differential equation in $V(\psi
).$ For this to happen, the coefficient of $V^{\prime }$ as well as the
function $N$ should be functions of $\psi $. That is 
\begin{subequations}
\label{psne}
\begin{eqnarray}
(\frac{\psi _{\lambda \mu }}{\psi _{\lambda }\psi _{\mu }})_{\lambda }\psi
_{\mu }-(\frac{\psi _{\lambda \mu }}{\psi _{\lambda }\psi _{\mu }})_{\mu
}\psi _{\lambda } &=&0  \label{pse} \\
N_{\lambda }\psi _{\mu }-N_{\mu }\psi _{\lambda } &=&0  \label{ne}
\end{eqnarray}%
Thus, for the present procedure to work, $\psi $ must satisfy simultaneously
the linear second-order equation (\ref{re1}) and the nonlinear third-order
equation (\ref{pse}).

We now begin with trying a solution of (\ref{re1}) in the form of one of the
functions $\psi _{n}$ and $\tilde{\psi}_{n}$ from (\ref{pss},\ref{psr}). It
turned out that only four functions $\psi _{1},\psi _{2},\tilde{\psi}_{1},%
\tilde{\psi}_{2}$ satisfy (\ref{pse}). Moreover, we have tried solutions in
the form of linear combinations 
\end{subequations}
\begin{equation}
\psi =\dsum\limits_{i=1}^{5}(J_{i}\psi _{i}+K_{i}\tilde{\psi}_{i})
\end{equation}%
This yielded the above four functions and two more functions $\psi _{1}\pm 2%
\tilde{\psi}_{1}=(\sqrt{\lambda }\pm \sqrt{\mu })^{2},$ so that the present
procedure can be applied for the six cases:

\begin{center}
\begin{tabular}{ll}
case & $\psi $ \\ 
17 & $2J\sqrt{\lambda \mu }$ \\ 
18 & $J(\lambda +\mu )$ \\ 
19 & $J(\lambda -\mu )^{2}$ \\ 
20 & $J\frac{\lambda +\mu }{\sqrt{\lambda \mu }}$ \\ 
21 & $J(\sqrt{\lambda }+\sqrt{\mu })^{2}$ \\ 
22 & $J(\sqrt{\lambda }-\sqrt{\mu })^{2}$%
\end{tabular}

Table II
\end{center}

Those cases will be discussed in the next section. For each case:

\begin{enumerate}
\item[a)] The function $\psi $ is substituted in (\ref{ne}) and separation
of variables is achieved after affecting a differential operator.

\item[b)] $F(\lambda ),G(\mu )$ are determined. It turned out that $F$ is
rational in case 17 and polynomial in all other cases and also $G(\mu )=-F$ $%
(\mu )$ in all cases except 19 and 20.

\item[c)] $F(\lambda ),G(\mu )$ are substituted in (\ref{Ne}) and the
expression $N$ is expressed as a function of $\psi $ only.

\item[d)] $N(\psi )$ is inserted in (\ref{Ve}), and the last is solved for $%
V(\psi ).$

\item[e)] The final form of $V$ is obtained by expressing $\psi $ again in
terms of $\lambda ,\mu .$
\end{enumerate}

Some of the indicated steps are quite cumbersome, even with the use of
computer algebra packages. For space saving most of those details are not
reflected here. We shall give only the data necessary to unambiguously
reproduce every case.

\subsubsection{The second procedure:}

The last procedure determined cases in which the potential $V$ is a solution
of a second order ODE (\ref{Ve}) depending linearly on one significant
arbitrary integration constant. The other being an immaterial additive
constant. This does not exclude the existence of other potentials not
containing this arbitrary constant. It may of course depend on the
parameters appearing in other functions $\psi ,F,G.$ For such cases $\psi $
is not required to satisfy (\ref{pse}-\ref{ne}). One may try expressing $%
V(\psi )$ in the form of a polynomial (say) 
\begin{equation}
V=\dsum\limits_{i=1}^{N}v_{i}\psi ^{i},
\end{equation}%
choose a solution $\psi $ of (\ref{re1}) and insert this expression in (\ref%
{re2a}) and try to find combinations of parameters that lead to separation
of the two unknown functions $F(\lambda ),G(\mu ).$ Obviously, the outcome
of this method will depend on the ability to perform necessary operations on
functions $V,\psi $ involving as much coefficients as possible.

We have made three trials:

\begin{enumerate}
\item[A)] For $\psi =\dsum\limits_{i=1}^{N}J_{i}\psi _{i}$ the separation
resulted in 12th-degree polynomials for $F$ and $G,$ but the final analysis
led to 6 cases

\begin{enumerate}
\item Two special versions of cases 18 and 19 of table II.

\item $\psi $ is a combination of $\psi _{1},\psi _{2},v_{2}=0,F,G$ are
cubic and $G(\mu )=-F(\mu ).$

\item $\psi $ is a combination of $\psi _{1},...,\psi _{3},v_{2}=0,F,G$ are
quadratic and $G(\mu )=-F(\mu ).$

\item $\psi $ is a combination of $\psi _{1},...,\psi _{4},v_{2}=0,F,G$ are
linear and $G(\mu )=-F(\mu ).$

\item $\psi $ is a combination of $\psi _{1},...,\psi _{5},v_{2}=0,F(\mu
)=1=-G(\mu ).$

It was noted that the presence of any of the few next $\psi _{i}$ with $ii>5$
led to the inconsistent result $G(\mu )=F(\mu )=0.$
\end{enumerate}

\item[B)] For $\psi =\dsum\limits_{i=1}^{5}J_{i}\tilde{\psi}_{i}$ we have
been lead to 5 cases:

\begin{enumerate}
\item A special version of case 17 of table II.

\item $\psi $ is a combination of $\tilde{\psi}_{1},\tilde{\psi}%
_{2},v_{1}=0,F,G$ are of 5th degree with two equal roots and $G(\mu )=-F(\mu
).$

\item $\psi $ is a combination of $\tilde{\psi}_{1},...,\tilde{\psi}%
_{3},v_{1}=0,F,G$ are of 5th degree with three equal roots and $G(\mu
)=-F(\mu ).$

\item $\psi $ is a combination of $\tilde{\psi}_{1},...,\tilde{\psi}%
_{4},v_{1}=0,F,G$ are of 5th degree with four equal roots and $G(\mu
)=-F(\mu ).$

\item $\psi $ is a combination of $\tilde{\psi}_{1},...,\tilde{\psi}%
_{5},v_{1}=0,F,G$ are of 5th degree with all five roots equal and $G(\mu
)=-F(\mu ).$
\end{enumerate}

\item[C)] For $\psi =\dsum\limits_{i=1}^{5}J_{i}\psi
_{i}(x-c_{i},y-c_{i}),v_{1}=0,F,G$ are of 5th degree and $G(\mu )=-F(\mu ),$
it was found that the only condition is that $\{c_{i}\}$ are roots of $F.$
\end{enumerate}

\section{\protect\bigskip The basic cases}

\subsection{\label{I}Case 17: The generating system with algebraic $\protect%
\psi $}

Here we try the first term of the sequence (\ref{psr}): 
\begin{equation}
\psi =2J\sqrt{\lambda \mu }  \label{ps1}
\end{equation}%
one is lead by equations (\ref{ne}) and then (\ref{Ve}) to rational
expression for $F$ 
\begin{equation}
F(\lambda )=a_{5}\lambda ^{5}+a_{4}\lambda ^{4}+a_{3}\lambda
^{3}+a_{2}\lambda ^{2}+a_{1}\lambda +a_{0}+\frac{b}{\lambda },\quad G(\mu
)=-F(\mu )  \label{rat}
\end{equation}%
The Lagrangian is given by%
\begin{eqnarray}
L &=&\frac{1}{2}(\lambda -\mu )[\frac{\dot{\lambda}^{2}}{F(\lambda )}-\frac{%
\dot{\mu}^{2}}{F(\mu )}]  \nonumber \\
&&-\frac{J\sqrt{-F(\lambda )F(\mu )}}{(\lambda -\mu )\sqrt{\lambda \mu }}[%
\frac{\lambda }{F(\lambda )}\dot{\lambda}-\frac{\mu }{F(\mu )}\dot{\mu}] 
\nonumber \\
&&+\frac{1}{2}a_{5}J^{2}\lambda \mu -\frac{bJ^{2}}{2\lambda ^{2}\mu ^{2}}+%
\frac{c}{\lambda \mu }  \label{LI}
\end{eqnarray}%
$a_{i},b,c$ being integration constants. The second integral is 
\begin{eqnarray}
I &=&\frac{1}{2}(\lambda -\mu )[\frac{\mu \dot{\lambda}^{2}}{F(\lambda )}-%
\frac{\lambda \dot{\mu}^{2}}{F(\mu )}]  \nonumber \\
&&+\frac{J\sqrt{-F(\lambda )F(\mu )}}{(\lambda -\mu )\sqrt{\lambda \mu }}[%
\frac{\lambda }{F(\lambda )}\dot{\lambda}+\frac{\mu }{F(\mu )}\dot{\mu}] 
\nonumber \\
&&+\frac{1}{2}J^{2}\{\frac{2b(\lambda +\mu )}{\lambda ^{2}\mu ^{2}}+\frac{%
a_{0}}{\lambda \mu }+\lambda \mu \lbrack a_{5}(\lambda +\mu )+a_{4}]\}+c%
\frac{\lambda +\mu }{\lambda \mu }
\end{eqnarray}%
The Gaussian curvature of the configuration space carrying (\ref{LI}) is%
\begin{equation}
\kappa =-\frac{1}{4}[a_{5}(3\lambda ^{2}+3\mu ^{2}+4\lambda \mu
)+2a_{4}(\lambda +\mu )+a_{3}-\frac{b}{\lambda ^{2}\mu ^{2}}]  \label{gcI}
\end{equation}

In order that the system described by the Lagrangian (\ref{LI}) was real, it
is necessary that the variable $\lambda $ varies on an interval on which $%
F(\lambda )\geq 0$ and $\mu $ on one of the intervals where $F(\mu )\leq 0.$
For the quadratic part of the kinetic energy to be positive definite the
factor $\lambda -\mu $ must be non-negative. The interval chosen for $%
\lambda $ should lie on the right of that for $\mu .$The number of working
combinations of intervals depends on the number of real roots of $F.$ Figure
4 shows possible choices of intervals for $\lambda $ and $\mu $ in the case
of five real distinct roots for $a_{5}>0.$ When $a_{5}<0$ the intervals are
interchanged and some of them are excluded to sustain the above conditions.%
\[
\FRAME{itbpF}{4.0603in}{2.5936in}{0in}{}{}{Figure}{\special{language
"Scientific Word";type "GRAPHIC";maintain-aspect-ratio TRUE;display
"USEDEF";valid_file "T";width 4.0603in;height 2.5936in;depth
0in;original-width 4.0101in;original-height 2.5521in;cropleft "0";croptop
"1";cropright "1";cropbottom "0";tempfilename
'JIK4W104.wmf';tempfile-properties "XPR";}}
\]

\begin{center}
Figure 4: Possible choices of intervals
\end{center}

The possible 9 combinations are listed in the following table

\begin{center}
$%
\begin{array}{cc}
a_{5}>0 & \left\{ 
\begin{tabular}{ll}
$\lambda $ & $\mu $ \\ 
\textit{I} & \textit{i,ii,iii} \\ 
\textit{II} & \textit{ii,iii} \\ 
\textit{III} & \textit{iii}%
\end{tabular}%
\right\} \\ 
a_{5}<0 & \left\{ 
\begin{tabular}{ll}
i\quad & II,III \\ 
ii & III%
\end{tabular}%
\right\}%
\end{array}%
$

Table III
\end{center}

The general picture changes when $F$ has equal roots or has pairs of complex
conjugate roots. Degenerate cases will be treated in the same way.

\bigskip In a next section we shall point out an important application of
the system I, Clebsch's case of motion of a solid in a liquid obtained as a
special case. Here we examine integrable motions on spaces of constant
Gaussian curvature. In fact, when $b=a_{5}=a_{4}=0$ the Gaussian curvature (%
\ref{gcI}) is constant $\kappa =-\frac{a_{3}}{4}$ and the Lagrangian takes
the form%
\begin{eqnarray}
L &=&\frac{1}{2}(\lambda -\mu )[\frac{\dot{\lambda}^{2}}{a_{3}\lambda
^{3}+a_{2}\lambda ^{2}+a_{1}\lambda +a_{0}}-\frac{\dot{\mu}^{2}}{a_{3}\mu
^{3}+a_{2}\mu ^{2}+a_{1}\mu +a_{0}}]  \nonumber \\
&&-J\frac{\sqrt{a_{3}\lambda ^{3}+a_{2}\lambda ^{2}+a_{1}\lambda +a_{0}}%
\sqrt{-(a_{3}\mu ^{3}+a_{2}\mu ^{2}+a_{1}\mu +a_{0})}}{(\lambda -\mu )\sqrt{%
\lambda \mu }}\times  \nonumber \\
&&\times \lbrack \frac{\lambda \dot{\lambda}}{a_{3}\lambda ^{3}+a_{2}\lambda
^{2}+a_{1}\lambda +a_{0}}-\frac{\mu \dot{\mu}}{a_{3}\mu ^{3}+a_{2}\mu
^{2}+a_{1}\mu +a_{0}}]  \nonumber \\
&&+\frac{h_{2}}{\lambda \mu }(+h)  \label{Lc1}
\end{eqnarray}%
For this system one may use the substitutions $\lambda =P_{1}(\xi ),\mu
=P_{1}(i\eta ),$ where $P_{1}$ is the Weierstrass elliptic function
generated by the inversion of the integral%
\begin{equation}
\xi =\int^{\lambda }\frac{d\lambda }{\sqrt{a_{3}\lambda ^{3}+a_{2}\lambda
^{2}+a_{1}\lambda +a_{0}}}
\end{equation}%
to transform to isometric coordinates. However, this procedure is not
transparent enough to isolate cases of real Lagrangian systems. Suitable
substitutions vary according to the number of real roots of the cubic
polynomial, their relative order and to the intervals on which $\lambda $
and $\mu $ vary. We consider somewhat further the case of three different
real roots $a,b,c$ $(a>b>c),$ say. Noting the symmetry of the structure of
the Lagrangian with respect to the variables $\lambda ,\mu ,$ and to
guarantee that the kinetic energy is positive definite, there is no loss in
generality in assuming that $\lambda -\mu \geq 0.$ The variables $\lambda
,\mu $ must also take their values on intervals, on which the cubic function
has different signs.

\begin{enumerate}
\item When $a_{3}>0$ we have three working combinations of intervals for
which the Lagrangian describes a motion on the pseudo sphere (or on the
hyperbolic plane $H^{2}$):

\begin{enumerate}
\item 
\begin{enumerate}
\item $\lambda \in \lbrack a,\infty ),\mu \in \lbrack b,a]$

\item $\lambda \in \lbrack a,\infty ),\mu \in (-\infty ,c]$

\item $\lambda \in \lbrack c,b],\mu \in (-\infty ,c]$
\end{enumerate}
\end{enumerate}

\item \label{pc}When $a_{3}<0$ we have only one working combination $\lambda
\in \lbrack b,a],\mu \in \lbrack c,b].$ This leads to a motion on the
sphere, which we will discuss in more detail in the next subsection.
\end{enumerate}

\subsubsection{\label{sphere}An integrable motion on a sphere}

In the case \ref{pc} of positive constant curvature, without loss of
generality one can set $a_{3}=-1.$ The Lagrangian and the integral can then
be put in the form%
\begin{eqnarray}
L &=&\frac{1}{2}(\lambda -\mu )[\frac{\dot{\lambda}^{2}}{(a-\lambda
)(\lambda -b)(\lambda -c)}+\frac{\dot{\mu}^{2}}{(a-\mu )(b-\mu )(\mu -c)}] 
\nonumber \\
&&-\frac{J}{4}\frac{\sqrt{(a-\lambda )(\lambda -b)(\lambda -c)}\sqrt{(a-\mu
)(b-\mu )(\mu -c)}}{(\lambda -\mu )\sqrt{\lambda \mu }}\times  \nonumber \\
&&\times \lbrack \frac{\lambda \dot{\lambda}}{(a-\lambda )(\lambda
-b)(\lambda -c)}-\frac{\mu \dot{\mu}}{(a-\mu )(b-\mu )(\mu -c)}]  \nonumber
\\
&&+\frac{h_{2}}{4\lambda \mu }(+\frac{h}{4})  \label{Lcc1}
\end{eqnarray}%
and 
\begin{eqnarray}
I &=&(\lambda -\mu )[\frac{\mu \dot{\lambda}^{2}}{(a-\lambda )(\lambda
-b)(\lambda -c)}+\frac{\lambda \dot{\mu}^{2}}{(a-\mu )(b-\mu )(\mu -c)}] 
\nonumber \\
&&-\frac{J}{\sqrt{\lambda \mu }}\sqrt{(a-\lambda )(\lambda -b)(\lambda -c)}%
\sqrt{(a-\mu )(b-\mu )(\mu -c)}\times  \nonumber \\
&&\times \lbrack \frac{\lambda \dot{\lambda}}{(a-\lambda )(\lambda
-b)(\lambda -c)}-\frac{\mu \dot{\mu}}{(a-\mu )(b-\mu )(\mu -c)}]  \nonumber
\\
&&-\frac{h_{2}(\lambda +\mu )}{\lambda \mu }+\frac{abcJ^{2}}{\lambda \mu }%
(+h)
\end{eqnarray}%
In the last system it is quite easy to recognize that $\lambda ,\mu $ are
the conical coordinates on the unit sphere. They are related to 3D Cartesian
coordinates through the relations 
\begin{equation}
x=\sqrt{\frac{(a-\lambda )(a-\mu )}{(a-b)(a-c)}},y=\sqrt{\frac{(\lambda
-b)(b-\mu )}{(a-b)(b-c)}},z=\sqrt{\frac{(\lambda -c)(\mu -c)}{(a-c)(b-c)}}
\end{equation}%
Note also that%
\begin{equation}
\frac{x^{2}}{a}+\frac{y^{2}}{b}+\frac{z^{2}}{c}=\frac{\lambda \mu }{abc}
\end{equation}%
and%
\begin{equation}
(b+c)x^{2}+(c+a)y^{2}+(a+b)z^{2}=\lambda +\mu
\end{equation}%
so that $\lambda ,\mu $ are the solutions of the quadratic equation 
\begin{equation}
\zeta ^{2}-[(b+c)x^{2}+(c+a)y^{2}+(a+b)z^{2}]\zeta +abc[\frac{x^{2}}{a}+%
\frac{y^{2}}{b}+\frac{z^{2}}{c}]=0
\end{equation}%
The potential term in the Lagrangian (\ref{Lcc1}) 
\begin{equation}
V_{0}=-\frac{h_{2}}{4\lambda \mu }=-\frac{h_{2}}{4abc(\frac{x^{2}}{a}+\frac{%
y^{2}}{b}+\frac{z^{2}}{c})}
\end{equation}%
is a separable potential. The Lagrangian (\ref{Lcc1}) thus generalizes the
separable system by the introduction of the gyroscopic terms. Comparing the
Lagrangian (\ref{Lcc1}) to the results of our paper \cite{ysph} we conclude
that the first gives a new integrable case of the motion of a particle on a
sphere or, alternatively, a new integrable case in rigid body dynamics under
forces with scalar and vector potential.

\subsubsection{Systems on a flat manifold}

The condition of vanishing Gaussian curvature $\kappa =0$ leads to the plane
case%
\begin{eqnarray}
V &=&\frac{C\alpha \left( \alpha -a\right) }{\left( a-\alpha \right)
x^{2}-\alpha y^{2}+\alpha a\left( \alpha -a\right) }  \nonumber \\
\Omega &=&\frac{1}{2}C_{1}\left( a-\alpha \right) [\frac{\alpha }{\left(
a-\alpha \right) x^{2}-\alpha y^{2}+\alpha a\left( \alpha -a\right) }]^{3/2}
\end{eqnarray}%
discussed firstly in section 3.4 of our work \cite{ypl}. A case presented in 
\cite{mcsw-win} can be easily obtained as a limiting case of this system. An
equivalent system was discussed in \cite{puc-ros} where one more limiting
case is pointed out.

\subsection{\label{II}Case 17: $\protect\psi $ is a linear function}

The choice $\psi =J(\lambda +\mu )$ leads to the expressions%
\begin{equation}
F(\lambda )=a_{6}\lambda ^{6}+a_{5}\lambda ^{5}+a_{4}\lambda
^{4}+a_{3}\lambda ^{3}+a_{2}\lambda ^{2}+a_{1}\lambda +a_{0},\quad G(\mu
)=-F(\mu )  \label{FII}
\end{equation}%
\begin{eqnarray}
L &=&\frac{1}{2}(\lambda -\mu )[\frac{\dot{\lambda}^{2}}{F(\lambda )}-\frac{%
\dot{\mu}^{2}}{F(\mu )}]-\frac{J\sqrt{-F(\lambda )F(\mu )}}{\lambda -\mu }[%
\frac{\dot{\lambda}}{F(\lambda )}-\frac{\dot{\mu}}{F(\mu )}]  \nonumber \\
&&+\frac{1}{2}J^{2}[a_{6}(\lambda +\mu )^{3}+a_{5}(\lambda +\mu
)^{2}]-K_{1}(\lambda +\mu )
\end{eqnarray}%
and the integral%
\begin{eqnarray}
I &=&\frac{1}{2}(\lambda -\mu )[\frac{\lambda \dot{\lambda}^{2}}{F(\lambda )}%
-\frac{\mu \dot{\mu}^{2}}{F(\mu )}]  \nonumber \\
&&+J\sqrt{-F(\lambda )F(\mu )}[\frac{\dot{\lambda}}{F(\lambda )}+\frac{\dot{%
\mu}}{F(\mu )}]  \nonumber \\
&&+\frac{1}{2}J^{2}\{(\lambda +\mu )[a_{6}(\lambda +\mu )(\lambda ^{2}+\mu
^{2}-\lambda \mu )+a_{5}(\lambda ^{2}+\mu ^{2})]  \nonumber \\
&&\qquad \qquad +a_{4}(\lambda +\mu )^{2}+a_{3}(\lambda +\mu
)\}-2K_{1}\lambda \mu
\end{eqnarray}%
\begin{eqnarray}
\kappa =-\frac{1}{4} &&[2a_{6}(\lambda +\mu )(2\lambda ^{2}+2\mu
^{2}+\lambda \mu )\qquad  \nonumber \\
&&+a_{5}(3\lambda ^{2}+3\mu ^{2}+4\lambda \mu )+2a_{4}(\lambda +\mu )+a_{3}]
\end{eqnarray}

\subsubsection{\label{cleb}The second case of Clebsch of motion of a solid
in a liquid}

Under the restrictions $a_{6}=a_{5}=a_{4}=0,$ the Gaussian curvature becomes 
$\kappa =-\frac{a_{3}}{4}.$ In order that the configuration manifold was an
ordinary sphere one may set $a_{3}=-1$ (say). The cubic polynomial must have
three different real roots $\,a,b,c(a>b>c)$ and the variables $\lambda ,\mu $
should be confined to the intervals $[b,a]$ and $[c,b],$ respectively, so
that $\lambda -\mu \geq 0.$ We obtain%
\begin{eqnarray}
L &=&\frac{1}{2}(\lambda -\mu )[\frac{\dot{\lambda}^{2}}{(a-\lambda
)(\lambda -b)(\lambda -c)}+\frac{\dot{\mu}^{2}}{(a-\mu )(b-\mu )(\mu -c)}] 
\nonumber \\
&&-J\frac{\sqrt{(a-\lambda )(\lambda -b)(\lambda -c)}\sqrt{(a-\mu )(b-\mu
)(\mu -c)}}{(\lambda -\mu )}\times  \nonumber \\
&&\times \lbrack \frac{\dot{\lambda}}{(a-\lambda )(\lambda -b)(\lambda -c)}+%
\frac{\dot{\mu}}{(a-\mu )(b-\mu )(\mu -c)}]  \nonumber \\
&&+h_{1}(\lambda +\mu )
\end{eqnarray}%
Comparing this to formulas of subsection (\ref{sphere}) above, we find that
the potential is%
\begin{eqnarray}
V &=&h_{1}(\lambda +\mu )  \nonumber \\
&=&h_{1}[(b+c)x^{2}+(c+a)y^{2}+(a+b)z^{2}]  \nonumber \\
&=&const-h_{1}(ax^{2}+by^{2}+cz^{2})
\end{eqnarray}%
Comparing potential and gyroscopic terms to the list of integrable motions
on a sphere in \cite{ysph}, one can show that we are dealing with the famous
case of motion of a rigid body in a liquid named as the second case of
Clebsch, characterized by a spherically symmetric inertia matrix. We shall
not reproduce the integral in this case, which is well known.

\subsection{\label{III}Case 19: $\protect\psi =J(\protect\lambda -\protect%
\mu )^{2}$}

For this family of systems

\begin{eqnarray}
F(\lambda ) &=&a_{3}\lambda ^{3}+a_{2}\lambda ^{2}+a_{1}\lambda +a_{0}, 
\nonumber \\
G(\mu ) &=&-a_{3}\mu ^{3}+b_{2}\mu ^{2}+b_{1}\mu +b_{0}
\end{eqnarray}%
\begin{eqnarray}
L &=&\frac{1}{2}(\lambda -\mu )[\frac{\dot{\lambda}^{2}}{F(\lambda )}+\frac{%
\dot{\mu}^{2}}{G(\mu )}]+J\sqrt{\frac{G(\mu )}{F(\lambda )}}\dot{\lambda}-J%
\sqrt{\frac{F(\lambda )}{G(\mu )}}\dot{\mu}  \nonumber \\
&&+\frac{1}{2}J^{2}[a_{3}(\lambda -\mu )^{2}+(a_{2}+b_{2})(\lambda -\mu )]-%
\frac{K_{1}}{(\lambda -\mu )}
\end{eqnarray}%
\begin{eqnarray*}
I &=&\frac{1}{2}(\lambda -\mu )[\frac{\mu \dot{\lambda}^{2}}{F(\lambda )}+%
\frac{\lambda \dot{\mu}^{2}}{G(\mu )}]-J(\lambda -\mu )(\sqrt{\frac{G(\mu )}{%
F(\lambda )}}\dot{\lambda}+\sqrt{\frac{F(\lambda )}{G(\mu )}}\dot{\mu}) \\
&&+\frac{1}{2}J^{2}(\lambda -\mu )[a_{3}(\lambda ^{2}-\mu ^{2})+a_{2}\lambda
+b_{2}\mu +a_{1}+b_{1}]+\frac{1}{2}K_{1}\frac{\lambda +\mu }{\lambda -\mu }
\end{eqnarray*}%
and\bigskip 
\begin{equation}
\kappa =\frac{1}{4}\left[ -a_{3}+\frac{2(a_{2}+b_{2})\lambda \mu
+(a_{1}+b_{1})(\lambda +\mu )+2(a_{0}+b_{0})}{(\lambda -\mu )^{3}}\right]
\end{equation}

\subsection{\label{IV}Case 20: $\protect\psi =J\frac{\protect\lambda +%
\protect\mu }{\protect\sqrt{\protect\lambda \protect\mu }}$}

\bigskip For this case we get%
\begin{equation}
F(\lambda )=\lambda ^{2}(a_{3}\lambda ^{3}+a_{2}\lambda ^{2}+a_{1}\lambda
+a_{0}),\quad G(\mu )=\mu ^{2}(b_{3}\mu ^{3}+b_{2}\mu ^{2}+b_{1}\mu -a_{0})
\end{equation}%
Note that $F(0)+G(0)=0.$ After some tedious calculations we obtain as final
result of (\ref{Ve})%
\begin{equation}
V(\psi )=\frac{16A\psi -(a_{3}+b_{3})\psi ^{3}}{16\sqrt{\psi ^{2}-4J^{2}}}-%
\frac{1}{16}(a_{3}-b_{3})\psi ^{2}
\end{equation}%
The Lagrangian of the present system has the form%
\begin{eqnarray}
L &=&\frac{1}{2}(\lambda -\mu )[\frac{\dot{\lambda}^{2}}{F(\lambda )}+\frac{%
\dot{\mu}^{2}}{G(\mu )}]  \nonumber \\
&&+\frac{1}{2}J\sqrt{\frac{F(\lambda )G(\mu )}{\lambda ^{3}\mu ^{3}}}[\frac{%
\lambda \dot{\lambda}}{F(\lambda )}-\frac{\mu \dot{\mu}}{F(\mu )}]  \nonumber
\\
&&-A\frac{\lambda +\mu }{\lambda -\mu }+J^{2}\frac{(\lambda +\mu )^{2}}{%
8(\lambda -\mu )}(\frac{a_{3}}{\mu }+\frac{b_{3}}{\lambda })\qquad (+K)
\label{n1}
\end{eqnarray}%
This Lagrangian admits the integral%
\begin{eqnarray}
I &=&\frac{1}{2}(\lambda -\mu )[\frac{\mu \dot{\lambda}^{2}}{F(\lambda )}+%
\frac{\lambda \dot{\mu}^{2}}{G(\mu )}]  \nonumber \\
&&-\frac{1}{2}J(\lambda -\mu )\sqrt{\frac{F(\lambda )G(\mu )}{\lambda
^{3}\mu ^{3}}}[\frac{\lambda \dot{\lambda}}{F(\lambda )}+\frac{\mu \dot{\mu}%
}{F(\mu )}]  \nonumber \\
&&+A\frac{2\lambda \mu -a(\lambda +\mu )}{\lambda -\mu }  \nonumber \\
&&+\frac{J^{2}}{8}\{a_{2}\frac{\lambda }{\mu }-b_{2}\frac{\mu }{\lambda }+%
\frac{\lambda +\mu }{\lambda -\mu }[a_{3}\frac{\lambda }{\mu }(\lambda -3\mu
)-b_{3}\frac{\mu }{\lambda }(3\lambda -\mu )]\}
\end{eqnarray}

The system (\ref{n1}) is completely new. The fifth degree polynomials $F,G$
have an identical double zero root but the other three are independent for
each polynomial. The Gaussian curvature corresponding to (\ref{n1}) can be
put in the form%
\begin{eqnarray}
\kappa &=&-\frac{1}{4}\left[ a_{3}(3\lambda ^{2}+3\mu ^{2}+4\lambda \mu
)+2a_{4}(\lambda +\mu )+a_{3}\right]  \nonumber \\
&&+\frac{(a_{3}+b_{3})y^{4}(5\lambda -3\mu )+(a_{2}+b_{2})y^{3}(2\lambda
-\mu )+(a_{1}+b_{1})y^{2}(3\lambda -\mu )}{4(\lambda -\mu )^{3}}
\end{eqnarray}

\subsection{\label{V}Cases 21 and 22: $\protect\psi =J(\protect\sqrt{\protect%
\lambda }\pm \protect\sqrt{\protect\mu })^{2}$}

\bigskip As in the previous section, for the present two cases we obtain in
final form%
\begin{eqnarray}
L &=&\frac{1}{2}(\lambda -\mu )[\frac{\dot{\lambda}^{2}}{F(\lambda )}-\frac{%
\dot{\mu}^{2}}{F(\mu )}]  \nonumber \\
&&-J\frac{(\sqrt{\lambda }\pm \sqrt{\mu })}{(\lambda -\mu )}\sqrt{\frac{%
-F(\lambda )F(\mu )}{\lambda \mu }}[\frac{\sqrt{\lambda }\dot{\lambda}}{%
F(\lambda )}+\frac{\sqrt{\mu }\dot{\mu}}{F(\mu )}]  \nonumber \\
&&-\frac{A}{\sqrt{\lambda }\pm \sqrt{\mu }}+\frac{1}{2}J^{2}a_{4}(\sqrt{%
\lambda }\pm \sqrt{\mu })^{2}
\end{eqnarray}%
\begin{equation}
F(\lambda )=\lambda (a_{3}\lambda ^{3}+a_{2}\lambda ^{2}+a_{1}\lambda +a_{0})
\end{equation}%
This system will have the integral%
\begin{eqnarray}
I &=&\frac{1}{2}(\lambda -\mu )[\frac{\mu \dot{\lambda}^{2}}{F(\lambda )}-%
\frac{\lambda \dot{\mu}^{2}}{F(\mu )}]  \nonumber \\
&&-J(\sqrt{\lambda }\pm \sqrt{\mu })\sqrt{\frac{-F(\lambda )F(\mu )}{\lambda
\mu }}[\frac{\sqrt{\lambda }\dot{\lambda}}{F(\lambda )}-\frac{\sqrt{\mu }%
\dot{\mu}}{F(\mu )}]  \nonumber \\
&&-A\frac{\sqrt{\lambda \mu }}{\sqrt{\lambda }\pm \sqrt{\mu }}+\frac{1}{2}%
J^{2}[a_{3}(\lambda +\mu )+a_{2}](\sqrt{\lambda }\pm \sqrt{\mu })^{2}
\end{eqnarray}%
The two systems just obtained are different, in the sense that the
difference in sign is significant. In fact, one can see that as $\lambda
,\mu $ approach one and the same root of $F$ (different from $0$) the
potential term $\frac{A}{\sqrt{\lambda }\pm \sqrt{\mu }}$ remains finite for
the positive sign but becomes infinite with the negative sign. The present
two families of systems are new. Both families share the same configuration
space. The Lagrangians can be explicitly in terms of global isometric
variables $\xi ,\eta $ using elliptic functions of complementary moduli. The
Gaussian curvature of the configuration space can be calculated as 
\begin{equation}
\kappa =-\frac{1}{4}\left[ 2a_{3}(\lambda +\mu )+a_{2}\right]
\end{equation}%
When $a_{3}=0$ this space becomes of constant curvature.

\section{Variations of cases 17 and 18}

It was noted in \cite{y92} that if we choose $F(\lambda )$ as the polynomial
part of (\ref{rat}), i.e.%
\begin{equation}
F(\lambda )=a_{5}\lambda ^{5}+a_{4}\lambda ^{4}+a_{3}\lambda
^{3}+a_{2}\lambda ^{2}+a_{1}\lambda +a_{0}  \label{F5}
\end{equation}%
then one can take $\psi $ as a linear combination of five terms of the type (%
\ref{ps1})%
\begin{equation}
\dsum\limits_{j}J_{j}\sqrt{(\lambda -\nu _{j})(\mu -\nu _{j})}  \label{ps5}
\end{equation}%
provided $\{$ $\nu _{j},j=1,...,5\}$ are the roots of the equation $F(\nu
)=0.$ Here $J_{j}$ are arbitrary constants which may be chosen real or
complex in such a way that $\psi $ is real. The system constructed in this
way involves 11 arbitrary parameters $\{a_{0},...,a_{5},J_{1},...,J_{5}\}$.
When some of the roots $\{\nu _{j}\}$ are equal, some of the terms in (\ref%
{ps5}) becomes dependent of the others and the number of significant
parameters $\{J_{j}\}$ is reduced by the number of extra repetition of
roots. It turned out that the coalescence of roots can be made in such a way
to preserve the total number of free parameters in the system.

\begin{enumerate}
\item The first procedure was indicated in \cite{y92}. A root $\nu $
repeated $(k+1)$ times leads to the replacement of $k$ terms in (\ref{ps5})
by terms of the form%
\[
\sum\limits_{i=1}^{k}J_{i}^{\prime }\frac{d^{i}}{d\nu ^{i}}\sqrt{(\lambda
-\nu )(\mu -\nu )} 
\]%
so that we have the sequence of terms that can appear at possible
coalescence combinations are obtained from (\ref{psr}) by a shift $\nu $ in
both variables.

\item When the leading coefficient $a_{5}$ of the polynomial (\ref{F5})
vanishes one of the roots goes to infinity. The same happens with every
vanishing leading coefficient. It turned out that for every infinite root
one can associate one of the terms in $\psi $ of the sequence obtained by
expanding the function $\sqrt{(\lambda -\nu )(\mu -\nu )}/\nu $ in powers of 
$\frac{1}{\nu }.$ This gives again the sequence of expressions (\ref{pss}).
Terms of this sequence appear in $\psi $ according to the number of infinite
roots.
\end{enumerate}

In the following subsections we provide information, sufficient for clear
and easy characterization of all the possible $19$ cases with different
combinations of coalescence of roots of the above two types. For every case
we give only the forms of $F,\psi $ and the potential function $V(\psi ).$
The Lagrangian is given by (\ref{Lrt}). The integral can be constructed in a
way similar to that followed in the above cases.

We shall identify each case by an ordered triplet expressing, respectively,
the number of equal finite roots, the number of finite roots which are not
equal to any others and the number of infinite roots. The first number may
consist of two parts, when there are two sets of equal finite roots.
Possible numbers are $(2,2)$ and $(2,3)$ only. In this classification the
primitive case of five distinct roots will carry the triplet $[0,5,0].$ The
case of five infinite roots $[0,0,5],$ the same as $F(\lambda )=1,$ leads to
a case of pseudo-Riemannian metric (on which the kinetic energy is not
positive definite) and we shall keep such cases out of the main list and
list them separately. The 19 cases follow in the next two subsections

\subsection{Classification of the resulting cases}

With the appropriate choice of the intervals for $\lambda ,\mu ,$ the
following 19 cases admit interpretation as motions on Riemannian or
pseudo-Riemannian manifolds:

\newpage

\begin{center}
\begin{tabular}{|l|l|l|l|l|}
\hline
& Class & $F(\lambda )$ & $\psi $ & $V$ \\ \hline
23 & $[0,5,0]$ & $\sum\limits_{j=0}^{5}a_{i}\lambda ^{i}$ & $%
2\sum\limits_{j=1}^{5}J_{j}\sqrt{(\lambda -\nu _{j})(\mu -\nu _{j})}$ & $-%
\frac{a_{5}}{8}\psi ^{2}$ \\ \hline
24 & $[2,3,0]$ & $a_{5}\prod\limits_{1}^{3}(\lambda -\nu _{i})(\lambda -\nu
_{4})^{2}$ & $%
\begin{tabular}{l}
$2\sum\limits_{j=1}^{4}J_{j}\sqrt{(\lambda -\nu _{j})(\mu -\nu _{j})}$ \\ 
$+J_{5}\frac{\lambda +\mu -2\nu _{4}}{\sqrt{(\lambda -\nu _{4})(\mu -\nu
_{4})}}$%
\end{tabular}%
$ & $-\frac{a_{5}}{8}\psi ^{2}$ \\ \hline
25 & $[3,2,0]$ & $%
\begin{tabular}{l}
$a_{5}(\lambda -\nu _{1})(\lambda -\nu _{2})\times $ \\ 
$\times (\lambda -\nu )^{3}$%
\end{tabular}%
$ & 
\begin{tabular}{l}
$2\sum\limits_{j=1}^{3}J_{j}\sqrt{(\lambda -\nu _{j})(\mu -\nu _{j})}$ \\ 
$+J_{4}\frac{\lambda +\mu -2\nu _{3}}{\sqrt{(\lambda -\nu _{3})(\mu -\nu
_{3})}}+J_{5}\frac{(\lambda -\mu )^{2}}{[(\lambda -\nu _{3})(\mu -\nu
_{3})]^{3/2}}$%
\end{tabular}
& $-\frac{a_{5}}{8}\psi ^{2}$ \\ \hline
26 & $[4,1,0]$ & $a_{5}(\lambda -\nu _{1})(\lambda -\nu )^{4}$ & 
\begin{tabular}{l}
$2J_{1}\sqrt{(\lambda -\nu _{1})(\mu -\nu _{1})}$ \\ 
$+2J_{2}\sqrt{(\lambda -\nu )(\mu -\nu )}+J_{3}\frac{\lambda +\mu -2\nu }{%
\sqrt{(\lambda -\nu )(\mu -\nu )}}$ \\ 
$+J_{4}\frac{(\lambda -\mu )^{2}}{[(\lambda -\nu )(\mu -\nu )]^{3/2}}+J_{5}%
\frac{(\lambda -\mu )^{2}(\lambda +\mu -2\nu )}{[(\lambda -\nu )(\mu -\nu
)]^{5/2}}$%
\end{tabular}
& $-\frac{a_{5}}{8}\psi ^{2}$ \\ \hline
27 & $[(2,2),1,0]$ & $%
\begin{tabular}{l}
$a_{5}(\lambda -\nu _{1})^{2}\times $ \\ 
$\times (\lambda -\nu _{2})^{2}(\lambda -\nu _{3})$%
\end{tabular}%
$ & $%
\begin{tabular}{l}
$2\sum\limits_{j=1}^{2}[J_{j}\sqrt{(\lambda -\nu _{j})(\mu -\nu _{j})}$ \\ 
$\qquad +J_{j+2}\frac{\lambda +\mu -2\nu _{j}}{\sqrt{(\lambda -\nu _{j})(\mu
-\nu _{j})}}]$ \\ 
$+2J_{5}\sqrt{(\lambda -\nu _{3})(\mu -\nu _{3})}$%
\end{tabular}%
$ & $-\frac{a_{5}}{8}\psi ^{2}$ \\ \hline
28 & $[(2,3),0,0]$ & $a_{5}(\lambda -\nu _{1})^{2}(\lambda -\nu _{2})^{3}$ & 
$%
\begin{tabular}{l}
$2\sum\limits_{j=1}^{2}[J_{j}\sqrt{(\lambda -\nu _{j})(\mu -\nu _{j})}$ \\ 
$\qquad +J_{j+2}\frac{\lambda +\mu -2\nu _{j}}{\sqrt{(\lambda -\nu _{j})(\mu
-\nu _{j})}}]$ \\ 
$+2J_{5}\frac{(\lambda -\mu )^{2}}{[(\lambda -\nu _{2})(\mu -\nu _{2})]^{3/2}%
}$%
\end{tabular}%
$ & $-\frac{a_{5}}{8}\psi ^{2}$ \\ \hline
29 & $[5,0,0]$ & $a_{5}(\lambda -\nu _{1})^{5}$ & $%
\begin{tabular}{l}
$J_{1}\sqrt{\lambda \mu }+J_{2}\frac{\lambda +\mu }{\sqrt{\lambda \mu }}%
+J_{3}\frac{(\lambda -\mu )^{2}}{(\lambda \mu )^{3/2}}$ \\ 
$+J_{4}\frac{(\lambda -\mu )^{2}(\lambda +\mu )}{(\lambda \mu )^{5/2}}+J_{5}%
\frac{(\lambda -\mu )^{2}(5\lambda ^{2}+5\mu ^{2}+6\lambda \mu )}{(\lambda
\mu )^{7/2}}$%
\end{tabular}%
$ & $-\frac{a_{5}}{8}\psi ^{2}$ \\ \hline
30 & $[0,4,1]$ & 
\begin{tabular}{l}
$a_{4}\lambda ^{4}+a_{3}\lambda ^{3}+a_{2}\lambda ^{2}$ \\ 
$+a_{1}\lambda +a_{0}$%
\end{tabular}
& $2\sum\limits_{j=1}^{4}J_{j}\sqrt{(\lambda -\nu _{j})(\mu -\nu _{j})}%
+J_{5}(\lambda +\mu )$ & $-\frac{a_{4}}{2}J_{5}\psi $ \\ \hline
31 & $[2,2,1]$ & $%
\begin{tabular}{l}
$a_{4}(\lambda -\nu _{1})(\lambda -\nu _{2})\times $ \\ 
$\times (\lambda -\nu _{3})^{2}$%
\end{tabular}%
$ & $%
\begin{tabular}{l}
$2\sum\limits_{j=1}^{3}J_{j}\sqrt{(\lambda -\nu _{j})(\mu -\nu _{j})}$ \\ 
$-J_{4}\frac{\lambda +\mu -2\nu _{3}}{\sqrt{(\lambda -\nu _{3})(\mu -\nu
_{3})}}+J_{5}(\lambda +\mu )$%
\end{tabular}%
$ & $-\frac{a_{4}}{2}J_{5}\psi $ \\ \hline
32 & $[3,1,1]$ & $a_{4}(\lambda -\nu _{1})(\lambda -\nu )^{3}$ & $%
\begin{tabular}{l}
$2J_{1}\sqrt{(\lambda -\nu )(\mu -\nu )}+J_{2}\frac{\lambda +\mu -2\nu }{%
\sqrt{(\lambda -\nu )(\mu -\nu )}}$ \\ 
$+J_{3}\frac{(\lambda -\mu )^{2}}{[(\lambda -\nu )(\mu -\nu )]^{3/2}}$ \\ 
$+2J_{4}\sqrt{(\lambda -\nu _{1})(\mu -\nu _{1})}+J_{5}(\lambda +\mu )$%
\end{tabular}%
$ & $-\frac{a_{4}}{2}J_{5}\psi $ \\ \hline
33 & $[0,3,2]$ & $%
\begin{tabular}{l}
$a_{3}\lambda ^{3}+a_{2}\lambda ^{2}$ \\ 
$+a_{1}\lambda +a_{0}$%
\end{tabular}%
$ & $%
\begin{tabular}{l}
$2\sum\limits_{j=1}^{3}J_{j}\sqrt{(\lambda -\nu _{j})(\mu -\nu _{j})}%
+J_{4}(\lambda +\mu )$ \\ 
$+J_{5}(\lambda -\mu )^{2}$%
\end{tabular}%
$ & $-2a_{3}J_{5}\psi $ \\ \hline
\end{tabular}

\bigskip 
\begin{tabular}{|l|l|l|l|l|}
\hline
34 & $[2,1,2]$ & $a_{3}(\lambda -\nu _{1})(\lambda -\nu _{2})^{2}$ & $%
\begin{tabular}{l}
$2J_{1}\sqrt{(\lambda -\nu _{1})(\mu -\nu _{1})}$ \\ 
$+2J_{2}\sqrt{(\lambda -\nu _{2})(\mu -\nu _{2})}$ \\ 
$-J_{3}\frac{\lambda +\mu -2\nu _{2}}{\sqrt{(\lambda -\nu _{2})(\mu -\nu
_{2})}}$ \\ 
$+J_{4}(\lambda +\mu )+J_{5}(\lambda -\mu )^{2}$%
\end{tabular}%
$ & $-2a_{3}J_{5}\psi $ \\ \hline
35 & $[3,0,2]$ & $a_{3}\lambda ^{3}$ & $%
\begin{tabular}{l}
$iJ_{1}\sqrt{\lambda \mu }-4iJ_{2}\frac{\lambda +\mu }{\sqrt{\lambda \mu }}%
-16iJ_{3}\frac{(\lambda -\mu )^{2}}{(\lambda \mu )^{3/2}}$ \\ 
$+J_{4}(\lambda +\mu )+J_{5}(\lambda -\mu )^{2}$%
\end{tabular}%
$ & $-2a_{3}J_{5}\psi $ \\ \hline
36 & $[0,2,3]$ & $a_{2}\lambda ^{2}+a_{1}\lambda +a_{0}$ & $%
\begin{tabular}{l}
$J_{1}\sqrt{(\lambda -\nu _{1})(\mu -\nu _{1})}$ \\ 
$+J_{2}\sqrt{(\lambda -\nu _{2})(\mu -\nu _{2})}$ \\ 
$+J_{3}(\lambda +\mu )+J_{4}(\lambda -\mu )^{2}$ \\ 
$+J_{5}(\lambda +\mu )(\lambda -\mu )^{2}$%
\end{tabular}%
$ & $-4J_{2}a_{2}\psi $ \\ \hline
37 & $[0,1,4]$ & $a_{1}\lambda +a_{0}$ & 
\begin{tabular}{l}
$J_{1}\sqrt{-(\lambda +a_{0})(\mu +a_{0})}+J_{2}(\lambda +\mu )$ \\ 
$+(\lambda -\mu )^{2}[J_{3}+J_{4}(\lambda +\mu )$ \\ 
$\qquad \qquad +J_{5}(5\lambda ^{2}+6\lambda \mu +5\mu ^{2})]$%
\end{tabular}
& $-32a_{1}J_{5}\psi $ \\ \hline
38 & $[(2,2),0,1]$ & 
\begin{tabular}{l}
$a_{4}(\lambda -\nu _{1})^{2}$ \\ 
$\quad \times (\lambda -\nu _{2})^{2}$%
\end{tabular}
& 
\begin{tabular}{l}
$\sum\limits_{j=1}^{2}[J_{j}\sqrt{(\lambda -\nu _{j})(\mu -\nu _{j})}$ \\ 
$+J_{j+2}\frac{\lambda +\mu -2\nu _{j}}{\sqrt{(\lambda -\nu _{j})(\mu -\nu
_{j})}}]$ \\ 
$+J_{5}(\lambda +\mu )$%
\end{tabular}
& $-\frac{1}{2}a_{4}J_{5}\Psi $ \\ \hline
39 & $[4,0,1]$ & $a_{4}\lambda ^{4}$ & 
\begin{tabular}{l}
$J_{1}\sqrt{\lambda \mu }+J_{2}\frac{\lambda +\mu }{\sqrt{\lambda \mu }}%
+J_{3}\frac{(\lambda -\mu )^{2}}{(\lambda \mu )^{3/2}}$ \\ 
$+J_{4}\frac{(\lambda -\mu )^{2}(\lambda +\mu )}{(\lambda \mu )^{5/2}}%
+J_{5}(\lambda +\mu )$%
\end{tabular}
& $-\frac{1}{2}a_{4}J_{5}\Psi $ \\ \hline
40 & $[2,0,3]$ & $a_{2}\lambda ^{2}$ & 
\begin{tabular}{l}
$J_{1}\sqrt{\lambda \mu }+J_{2}\frac{\lambda +\mu }{\sqrt{\lambda \mu }}$ \\ 
$+J_{3}(\lambda +\mu )+J_{4}(\lambda -\mu )^{2}$ \\ 
$+J_{5}(\lambda -\mu )^{2}(\lambda +\mu )$%
\end{tabular}
& $-\frac{1}{4}a_{2}J_{5}\Psi $ \\ \hline
41 & $[0,0,5]$ & $1$ & 
\begin{tabular}{l}
$J_{1}(\lambda +\mu )+(\lambda -\mu )^{2}\times $ \\ 
$\times \lbrack J_{2}+J_{3}(\lambda +\mu )$ \\ 
$\quad +J_{4}(5\lambda ^{2}+6\lambda \mu +5\mu ^{2})$ \\ 
$\quad +J_{5}(\lambda +\mu )(7\lambda ^{2}+7\mu ^{2}+2\lambda \mu )]$%
\end{tabular}
& $64J_{5}\Psi $ \\ \hline
\end{tabular}

Table IV
\end{center}

\bigskip Among those cases one can easily recognize the three cases 33, 34
and 35 as cases of motion on spaces of constant curvature. The first of
those accommodates spheres and pseudo spheres. In fact, if the roots $\nu
_{i}$ are real and ordered such that $\nu _{1}>\nu _{2}>\nu _{3},$ then one
can choose for $\lambda $ and $\mu $ the intervals $\left[ \nu _{2},\nu _{1}%
\right] $ and $\left[ \nu _{3},\nu _{2}\right] ,$ respectively, and put $%
a_{3}=-1$ to be on a sphere. Similarly, if $a_{3}=1,\lambda \in \lbrack \nu
_{1},\infty )$ and $\mu \in (-\infty ,\nu _{3}]$ we have a pseudo-sphere. We
do not write Lagrangians in specific coordinates adapted to each of the
possible choices of intervals. In the other two cases only pseudo-sphere is
possible. The system 34 produces two possible Lagrangians. In case 35 the
only possible Lagrangian can be written as%
\begin{eqnarray}
L &=&\frac{1}{2}\frac{u^{2}+v^{2}}{u^{2}v^{2}}(\dot{u}^{2}+\dot{v}^{2}) 
\nonumber \\
&&+2\left\{ \left[ \frac{8J_{5}}{v^{3}}+\frac{J_{4}u^{2}}{v(u^{2}+v^{2})}%
\right] \dot{u}+\left[ \frac{8J_{5}}{u^{3}}-\frac{J_{4}v^{2}}{u(u^{2}+v^{2})}%
\right] \dot{v}\right\}  \nonumber \\
&&+128J_{5}^{2}\frac{(u^{2}+v^{2})^{2}}{u^{4}v^{4}}  \nonumber \\
&&+J_{5}\left[ 32J_{4}\frac{u^{2}-v^{2}}{u^{2}v^{2}}+J_{3}\frac{%
(u^{2}+v^{2})^{2}}{2uv}-4J_{2}\frac{(u^{2}-v^{2})}{uv}-32\frac{J_{1}}{uv}%
\right]
\end{eqnarray}%
Here we have omitted terms in the linear part of the Lagrangian, which have
no contribution to the gyroscopic function $\Omega .$

Cases 36 and 37 are Euclidean plane cases in coordinates of elliptic and
parabolic types. Case 36 is equivalent to the seven -parametric case given
in section 3.1 of \cite{ypl}. Case 37 can be shown to be representing in the 
$xy-$ plane a system with

\begin{equation}
V=Ax+By-a(x^{2}+y^{2})[C+bx+\frac{1}{2}a(5x^{2}+y^{2})],\quad \Omega =6ax+b
\end{equation}%
Trying to eliminate the parameter $B$ by a rotation of the axes would result
in new terms in both functions $\Omega ,V.$ Thus, this is a significant
generalization of a case found in \cite{y92} by introducing the parameter $B$
and freeing the parameter $A$ from a relation to $a.$ The integral for this
system is 
\begin{eqnarray}
I &=&a\left[ 2(x\dot{y}-y\dot{x})-\left( 4ax+b\right) \left(
x^{2}+y^{2}\right) \right] \left[ \dot{y}-C-bx-a\left( 3x^{2}+y^{2}\right) %
\right]  \nonumber \\
&&+A\left[ \dot{y}-bx-a\left( 3x^{2}+y^{2}\right) \right] -B\left[ \dot{x}%
+y\left( b+2ax\right) \right]
\end{eqnarray}

In the first fifteen cases of table IV the intervals on which the variables $%
\lambda ,\mu $ take their values could be chosen so that the configuration
manifold become real Riemannian. The remaining four cases, in which the
metric of the manifold has the single signature (1,-1) are listed as 38 to
41. We have just reduced the Lagrangians for those cases to real form by a
substitution $\psi =i\Psi ,$ where $\Psi $ is real. No effort is done to
reduce the cases obtained explicitly to the simplest form.

\section{More about interpretations}

In the previous sections we have constructed forty one different integrable
systems admitting a quadratic integral independent of Jacobi's integral. The
presence of several parameters in the structures of those systems gives a
wide possibility to adapt those parameters to fit concrete applications in
mechanics and physics. So far, we have identified some possible cases of
motion in the plane or developable surface, sphere or pseudo-sphere (or
hyperbolic plane) by analyzing the Gaussian curvature of the metric of the
system under consideration. The question arises, how to apply the above
results to a given problem in mechanics. The first and principal step is to
identify the metric of the configuration space of the given system with that
of one of the systems constructed in sections 3 - 7. This may require the
use of certain transformation of the generalized coordinates of the original
mechanical system to reach an isometric form with a coefficient $\Lambda $
of one of the types (\ref{ty1}) - (\ref{ty3}) and try to adapt the candidate
integrable system by imposing some restrictions on its parameters to fit it
to the given one. Once this step has succeeded, we are already dealing with
an integrable case of the given system and it only remains to characterize
the potential and electromagnetic forces in in terms of the original
coordinates of this system. If the original given mechanical system is
obtained from one of more than two degrees of freedom by means of reduction
in the sense of Routh, the cyclic constants of the motion would appear in
the linear and potential terms in a certain specified form and the
parameters of the candidate system may allow to accommodate those constants
either completely or partially. Extra-parameters not coming from cyclic
constant may then be considered for further explanation of their origins. In
the rest of this section we indicate certain applications of the above
results. For size considerations only some of them are considered in some
detail and others are pointed out.

\subsection{Application to rigid body dynamics}

As an example, we try here to find all cases of the above systems, relevant
to the dynamics of a rigid body moving about a fixed point. Reduction of the
most general problem of motion under the action of potential forces and
forces of electromagnetic type was reduced to the form involving a
Liouville-type metric in \cite{y92}. We give the final form of the
Lagrangian for the case of tri-axial ellipsoid of inertia. Let $\mathbf{I=}%
diag(A,B,C$ $)$ be the inertia matrix of the body in the principal axes of
inertia at the fixed point, $\mathbf{\omega =}(p,q,r)$ its angular velocity, 
$\mathbf{\gamma =}(\gamma _{1},$ $\gamma _{2},\gamma _{3})$ a unit vector
fixed in space and $V_{0}\left( \mathbf{\gamma }\right) ,\mathbf{m}\left( 
\mathbf{\gamma }\right) $ be the scalar and vector potentials of forces
acting on the body. The Lagrangian of the system has the form%
\begin{equation}
L=\frac{1}{2}\mathbf{\omega I\cdot \omega +m\cdot \omega -}V_{0}
\end{equation}%
The Euler-Poisson equations of motion can be written as%
\begin{equation}
\mathbf{\dot{\omega}I+\omega \times }(\mathbf{\omega I+M})=\mathbf{\gamma
\times }\frac{\partial V_{0}}{\partial \mathbf{\gamma }},\quad \mathbf{\dot{%
\gamma}+\omega \times \gamma =0}  \label{ep}
\end{equation}%
where 
\begin{equation}
\mathbf{M=}\frac{\partial }{\partial \mathbf{\gamma }}(\mathbf{m\cdot \gamma 
})-(\frac{\partial }{\partial \mathbf{\gamma }}\mathbf{\cdot m})\mathbf{%
\gamma }
\end{equation}%
This system admits the cyclic integral 
\begin{equation}
(\mathbf{\omega I+m})\mathbf{\cdot \gamma =}f  \label{ci}
\end{equation}%
corresponding to the angle of precession $\psi $ around the vector $\mathbf{%
\gamma }$ and allows reduction by the Routhian procedure to a system of two
degrees of freedom.

\subsubsection{\protect\bigskip Reduction in the case of a tri-axial body}

To describe the position of the current point on the configuration space of
the reduced system in this case, which is the Poisson sphere $\left\vert 
\mathbf{\gamma }\right\vert ^{2}=1$ we shall use the same variables $\lambda
,\mu $ as in (\cite{y92}). Those variables are equivalent to elliptic
coordinates on the ellipsoid of inertia with the standard metric. They are
solutions of the quadratic equation%
\begin{equation}
D\lambda ^{2}-N\lambda +ABC=0  \label{ellcoo}
\end{equation}%
\begin{eqnarray}
N &=&A(B+C)\gamma _{1}^{2}+B(C+A)\gamma _{2}^{2}+C(A+B)\gamma _{3}^{2}, 
\nonumber \\
D &=&A\gamma _{1}^{2}+B\gamma _{2}^{2}+C\gamma _{3}^{2}
\end{eqnarray}%
We get 
\begin{eqnarray}
R &=&\frac{1}{2}ABC(\lambda -\mu )[\frac{\dot{\lambda}^{2}}{\lambda
^{2}(A-\lambda )(\lambda -B)(\lambda -C)}+\frac{\dot{\mu}^{2}}{\mu
^{2}(A-\mu )(B-\mu )(\mu -C)}]  \nonumber \\
&&+\sqrt{ABC}[P\frac{\dot{\lambda}}{\lambda \sqrt{(A-\lambda )(\lambda
-B)(\lambda -C)}}-Q\frac{\dot{\mu}}{\mu \sqrt{(A-\mu )(B-\mu )(\mu -C)}}] 
\nonumber \\
&&-V  \label{R}
\end{eqnarray}%
where 
\begin{equation}
V=\frac{1}{4}[V_{0}+\frac{(f-m\cdot \gamma )^{2}}{2D}]\text{ }
\end{equation}%
We do not express $P,Q$ \ here explicitly as functions of $\lambda ,\mu $
and only compare the function 
\begin{eqnarray}
\Omega &=&\frac{1}{\sqrt{ABC}}[\mu \sqrt{(A-\mu )(B-\mu )(\mu -C)}\frac{%
\partial P}{\partial \mu }-\lambda \sqrt{(A-\lambda )(\lambda -B)(\lambda -C)%
}\frac{\partial Q}{\partial \lambda }]  \nonumber \\
&=&\frac{(\lambda -\mu )\sqrt{\lambda \mu }}{4ABC}\left\{ f\left[
A+B+C-2(\lambda +\mu )\right] +S\right\}
\end{eqnarray}%
in which $f$ is the cyclic constant and $S$ is a function depending on
vector-potential terms%
\begin{equation}
S=D\frac{\partial }{\partial \mathbf{\gamma }}\cdot \left[ \frac{1}{D}%
\mathbf{\gamma \times }(\mathbf{\gamma I\times m})\right]
\end{equation}

\subsubsection{\protect\bigskip Integrable cases of a tri-axial body}

We now look for cases of the constructed systems, which can accommodate the
metric in (\ref{R}). Note that here the functions $F$ and $G$ are
fifth-degree polynomials with a double zero root satisfying $G(\mu )=-F(\mu
),$ a condition common to most cases, but not all, of the previous sections.
We find that:

\begin{enumerate}
\item The case of \ref{I} under the restrictions $%
a_{1}=a_{0}=b=0.a_{2}=1,a_{3}=-(1/A+1/B+1/C),a_{4}=\frac{A+B+C}{ABC},a_{5}=%
\frac{1}{ABC}.$

\item Case 24 of table IV under the restriction $\nu _{4}=0.$

\item The case of \ref{II} under the same restrictions, but with $a_{6}$
instead of $b.$

\item The case of \ref{IV} under the restrictions $%
a_{0}=-b_{0}=1,a_{1}=-b_{1}=-(1/A+1/B+1/C),a_{2}=-b_{2}=\frac{A+B+C}{ABC}%
,a_{3}=-b_{3}=\frac{1}{ABC}.$
\end{enumerate}

It was shown in \cite{y92} that case 1 is a case of motion of a
multi-connected tri-axial solid in a liquid, known after Clebsch \cite{clb}
and case 2 is the case due originally to Steklov \cite{stk} in its
generalized form due to Rubanovsky \cite{rub} and involving circulations.
Case 3 was also pointed out as a new case. The reader is referred to \cite%
{y92} for full detail.

The last case turns out to be a new one different from known cases of rigid
body motion. For this case one can find 
\begin{eqnarray}
V_{0} &=&\frac{kN}{2ABC\sqrt{N^{2}-4ABCD}}  \nonumber \\
\mathbf{m} &=&\frac{K}{2ABC}(A(B+C)\gamma _{1},B(C+A)\gamma
_{2},C(A+B)\gamma _{3})  \nonumber \\
\mathbf{M} &=&-K(\frac{\gamma _{1}}{A},\frac{\gamma _{2}}{B},\frac{\gamma
_{3}}{C})
\end{eqnarray}%
where Equations (\ref{ep}) for this choice admit the conditional integral%
\begin{equation}
I=\frac{1}{2}(Ap^{2}+Bq^{2}+Cr^{2})-K(p\gamma _{1}+q\gamma _{2}+r\gamma
_{3})+\frac{k}{\sqrt{N^{2}-4ABCD}}
\end{equation}%
In fact, differentiating $I$ with respect to $t$ in virtue of (\ref{ep}) and
using the geometric and cyclic integrals%
\begin{eqnarray}
\gamma _{1}^{2}+\gamma _{2}^{2}+\gamma _{3}^{2} &=&1,  \nonumber \\
Ap\gamma _{1}+Bq\gamma _{2}+Cr\gamma _{3}+\frac{K}{2ABC}N &=&f  \label{ci2}
\end{eqnarray}%
we find%
\begin{equation}
\frac{dI}{dt}=4kfC(A-B)(B-C)(C-A)\gamma _{1}\gamma _{2}\gamma _{3}^{2}
\end{equation}%
and hence $\dot{I}$ vanishes identically in two cases:

\begin{enumerate}
\item When $k=0,$ and this reproduces the original Steklov case which is a
general integrable case for all initial conditions.

\item When $k\neq 0,f=0$ and this gives a new conditional case valid only on
the zero level of the cyclic integral (\ref{ci2}).
\end{enumerate}

\subsection{On time-reversible integrable problems in rigid body dynamics}

For the system (\ref{R}) to be time-reversible one must have $\mathbf{m=0}%
,f=0.$ The Lagrangian becomes

\begin{equation}
R=\frac{1}{2}ABC(\lambda -\mu )[\frac{\dot{\lambda}^{2}}{\lambda
^{2}(A-\lambda )(\lambda -B)(\lambda -C)}+\frac{\dot{\mu}^{2}}{\mu
^{2}(A-\mu )(B-\mu )(\mu -C)}]-\frac{1}{4}V_{0}
\end{equation}%
It is clear from the consideration of section 1 that the existence of a
quadratic integral is associated with the separation of variables. Comparing
the last system to (\ref{LS1}) and using (\ref{ellcoo}) we note that it
admits a quadratic integral if and only if 
\begin{equation}
V_{0}=D\frac{v_{1}(N+\sqrt{\beta })+v_{2}(N-\sqrt{\beta })}{\sqrt{\beta }}
\label{V0}
\end{equation}%
where $\beta =N^{2}-4ABCD$ and $v_{1},v_{2}$ are two arbitrary functions.
The explicit form of the integral in the Euler-Poisson variables is%
\begin{equation}
I=\frac{1}{2}(Ap^{2}+Bq^{2}+Cr^{2})+\frac{(N+\sqrt{\beta })v_{1}(N+\sqrt{%
\beta })+(N-\sqrt{\beta })v_{2}(N-\sqrt{\beta })}{\sqrt{\beta }}  \label{I0}
\end{equation}%
This can be verified by direct calculation. This result was obtained in \cite%
{yint86} and independently in equivalent form in \cite{bog}. In \cite{valent}
an infinite sequence of potentials $\left\{ U_{n},n=1...\infty \right\} $
that admit quadratic integral, using symbolic computation was found. It can
be easily shown that this sequence corresponds to the choices $\left\{
v_{1}(x)=-v_{2}(x)=x^{n}\right\} .$ This is exactly a case discussed
explicitly in \cite{bog}. Moreover, if in (\ref{V0}) we take $%
v_{1}(x)=-v_{2}(x)$ as a full power series we obtain the potential 
\begin{equation}
V_{0}=\dsum\limits_{1}^{\infty }A_{n}U_{n}
\end{equation}%
admitting the quadratic integral. This could be expected from the separation
property overlooked in \cite{valent}. It should be mentioned here that the
authors of \cite{valent} have also argued that the sequence they obtained is
different from another one, obtained earlier by Wojciechovski in 1985 \cite%
{woj}. However, the latter sequence as provided in \cite{valent} is not
separable for the tri-axial rigid body and is inconsistent with the
existence of a quadratic integral for rigid body dynamics.

\subsection{\protect\bigskip The problem of motion of a particle on a smooth
ellipsoid}

\bigskip Some mechanical systems do not have much importance in their own as
much as in other problems related to them, for which they serve as
geometrization. A typical example of a useful model in mechanics is the
problem of motion on a smooth fixed surface. The motion on a sphere is used
for modeling motion of a rigid body \cite{ysph}, in the study of B-phase of
the superfluid $^{3}He$, in the construction of certain wave solutions of
the Landau-Lifshitz non-linear equation (e.g. \cite{dub}) and in the
treatment of Dyson's fluid dynamical model of spinning gas clouds
maintaining ellipsoidal shape \cite{gaf1}. Recently, the motion of a
particle on a smooth tilted cone was found to be a generalization covering
the problem of motion of the swinging Atwood machine as a special case \cite%
{ycone}.

The ellipsoid is one of the favorite models. It is closely related to the
dynamics of rigid bodies. Let a particle of unit mass be moving on the
ellipsoid%
\begin{equation}
\frac{x^{2}}{a^{2}}+\frac{y^{2}}{b^{2}}+\frac{z^{2}}{c^{2}}=1
\end{equation}%
under the action of forces with potential $V$ and vector potential $\mathbf{l%
}$. The Lagrangian for this problem is%
\begin{equation}
L=\frac{1}{2}\mathbf{\dot{r}}^{2}+\mathbf{l}\cdot \mathbf{\dot{r}}-V
\end{equation}%
We shall only express $\mathbf{\dot{r}}^{2}$ in terms of elliptic
coordinates $u,v$ on the ellipsoid%
\begin{equation}
\frac{1}{4}(u-v)\left[ \frac{u\dot{u}^{2}}{(a^{2}-u)(u-b^{2})(u-c^{2})}-%
\frac{v\dot{v}^{2}}{(a^{2}-v)(v-b^{2})(v-c^{2})}\right]
\end{equation}%
Here the functions $F$ and $G$ are rational (cubic to linear). One can
immediately notice that case I (\ref{I}), in which $F$ is rational (degree 6
to one) gives a case of motion on the ellipsoid if coefficients are suitably
adapted. This case was found earlier in our work \cite{yell} by a different
method.

\subsection{Geometrical interpretation of systems of the type (\protect\ref%
{n})}

\bigskip Using Hamilton's principle in the form of Jacobi (e.g. \cite{pars}, 
\cite{gant}), one can derive the equations of the trajectories of motion of
the system described by the Lagrangian (\ref{nl}) from the variational
problem 
\begin{equation}
\delta \int \sqrt{2\Lambda (h-V)(d\xi ^{2}+d\eta ^{2})}+Pd\xi -Qd\eta =0
\end{equation}%
This means those trajectories are geodesics of the Randers metric 
\begin{equation}
ds=\sqrt{2\Lambda (h-V)(d\xi ^{2}+d\eta ^{2})}+Pd\xi -Qd\eta  \label{Rm}
\end{equation}

Randers metrics were first studied by physicist, G.Randers \cite{rand} from
the stand point of general relativity. Since then, many geometers have made
efforts in investigating the geometric properties of Randers metrics as an
important and rich class of Finsler metrics and to explore their prospective
applications (see e.g. \cite{rund} - \cite{boc}).

Every integrable mechanical system of the type (\ref{n}) corresponds to an
integrable geodesic flow of the Randers Metric (\ref{Rm}) admitting the
integral 
\begin{equation}
I=\Lambda (h-V)(\frac{d\xi }{dl})^{2}+\sqrt{2\Lambda (h-V)}(P\frac{d\xi }{dl}%
+Q\frac{d\eta }{dl})+R
\end{equation}%
in which $dl=\sqrt{d\xi ^{2}+d\eta ^{2}}.$

\subsection{\protect\bigskip The Fokker-Plank systems}

Stochastic dynamical models described by Fokker-Planck equations, in the
limit of weak noise, can be formally associated with Hamiltonian dynamical
systems. The Hamiltonian consists of a kinetic energy quadratic term and
terms linear in momenta , with zero scalar potential term (e.g. \cite{hiet}, 
\cite{gr}). The study of such Hamiltonian systems has received some
attention from both points of view of integrability \cite{rk} - \cite{kod}
and existence of invariant relations \cite{br}.

Gauge transformation of the Lagrangian was also used by Hietarinta \cite%
{hiet1} to construct integrable Fokker-Plank equations from known integrable
Hamiltonian systems. The Hamiltonian function corresponding to the
Lagrangian (\ref{nl}), with a gauge term $\frac{dZ(\xi ,\eta )}{dt}$, has
the form%
\begin{equation}
H=\frac{1}{2\Lambda }\left[ \left( p_{\xi }-P-\frac{dZ}{d\xi }\right)
^{2}+\left( p_{\eta }+Q-\frac{dZ}{d\eta }\right) ^{2}\right] +V=0
\end{equation}%
where $p_{\xi },p_{\eta }$ are the momenta conjugate to the coordinates $\xi
,\eta $ respectively.\ An integrable system (\ref{n}) generates a
Fokker-Plank system if there exists a function $Z(\xi ,\eta ),$ satisfying 
\begin{equation}
\frac{1}{2\Lambda }\left[ \left( P+\frac{dZ}{d\xi }\right) ^{2}+\left( Q-%
\frac{dZ}{d\eta }\right) ^{2}\right] +V=0
\end{equation}%
The results we obtained for this problem are promising and will be presented
elsewhere.

\section{Conclusion}

So far, we have made the most comprehensive analysis of the problem of
constructing 2D conservative Lagrangian systems involving terms linear in
the velocities, which admit a complementary quadratic integral. The problem
under consideration has proved unexpectedly rich. We have constructed and
tabulated 41 major several-parameter families of integrable systems of this
type. In most cases the parametrization of the system can be chosen, usually
in more than one way, so that the configuration manifold becomes Riemannian
or pseudo-Riemannian. The remaining few cases are real only on
pseudo-Riemannian manifolds.

Some of those families or their special cases or degenerations were listed
in our previous works but the majority are either listed for the first time
or involve more parameters in their structure. Usually, we give the
Lagrangian and the second invariant and occasionally the Gaussian curvature
of the configuration space, which we use for interpretation of simple
results as cases of motion on a plane or a space of constant curvature. In
application to the problem of motion of a rigid body under various
circumstances, all known integrable cases of the type under consideration
for this problem are restored and a new one is added. A systematic analysis
of the results of this paper is still needed to isolate all possible
subcases, which admit a preassigned physical or mechanical interpretation.
Many of them are tentatively candidates for the study of motion of a charged
particle under the action of spatially periodic electric and magnetic fields
on various 2D spaces.

It is worthy to say few words about the methods we used throughout the
paper, in order to make clear their limitations. The difficulty in solving
the problem stems from the fact that we deal with a linear PDE (\ref{re1})
for $\psi (\lambda ,\mu )$ coupled with a differential condition involving
nonlinearly the same function $\psi $ and linearly three functions of a
single variable each $F(\lambda ),G(\mu ),V(\psi ).$ This system is so
involved that we may not be able to tell at present how far we are from its
general solution. As we see from the last sections, $F$ and $G$ turned out
to be rational in one case and polynomial of degree less than six in all
others. We see from tables 4 and 5 how the degree of $F(\lambda )$ is
interlinked with the number of terms taken from the sequence of functions $%
\{\psi _{n},\tilde{\psi}_{n}\}.$ It seems unlikely that $F$ can be a
polynomial of degree six or higher. However, we are working on few cases in
which $F,G$ are algebraic functions. Using certain uniformizing
transformations, one can reduce $F,G$ to polynomial form, but this changes
the form of the metric coefficient $(\lambda -\mu ).$ This will be pursued
elsewhere.

In the more compact case considered in \ref{fp}, we deal with simultaneous
solutions of the linear PDE (\ref{re1}) and the nonlinear one (\ref{pse}) in
the same unknown function $\psi .$ It is unlikely that those equations admit
other solutions than the list of table II. However, we have no proof yet
that no other solutions exist, for example, as full infinite series of $%
\{\psi _{n},\tilde{\psi}_{n}\}.$

\textbf{Acknowledgement:} I am indebted to an anonymous referee, whose
remarks helped me reorganize this work and to reformulate the methods and
results in a much clearer way.

\end{document}